\documentclass[aps,prb,showpacs,floatfix,twocolumn,superscriptaddress]{revtex4-2}
\usepackage{graphicx}
\usepackage{dcolumn}
\usepackage{graphicx}
\usepackage{relsize}
\usepackage{xcolor}
\usepackage{cancel}
\usepackage{bm}
\usepackage{tikz}
\usepackage{siunitx}
\usepackage{tablefootnote}
\usepackage{physics}
\usepackage[section]{placeins}
\definecolor{darkblue}{rgb}{0,0,.65}
\usepackage{tikz}
\usepackage[%
 pdfauthor={Benedikt Placke},%
 pdfstartview=FitH,%
 breaklinks=true,%
 bookmarks=true,%
 colorlinks=true,%
 anchorcolor=black,%
 citecolor=red,
 filecolor=black,%
 menucolor=black,%
 urlcolor=blue,%
 linkcolor=r
 ed,%
]{hyperref}
\newcommand{\appref}[1]{\hyperref[#1]{App.~\ref*{#1}}}
\usepackage{physics}
\renewcommand\vec{\mathbf}
\usepackage[english]{babel}
\usepackage{bm}% mathbf math
\usepackage{amssymb}
\usepackage{color, soul}
\usepackage{amsmath}
\usepackage{amstext}
\usepackage{latexsym}
\usepackage{graphicx}% Include figure files
\usepackage{mathtools}
\newcommand{\rightTiltedTriangles}{
   \tikz[baseline=(current bounding box.center), scale=0.2]{
       \draw (0,0) -- (0.5,0.866) -- (1,0) -- cycle;
       \draw (0,0) -- (-0.5,-0.866) -- (-1,0) -- cycle;
   }
}

\newcommand{\leftTiltedTriangles}{
   \tikz[baseline=(current bounding box.center), scale=0.2]{
       \draw (0,0) -- (-0.5,0.866) -- (-1,0) -- cycle;
       \draw (0,0) -- (0.5,-0.866) -- (1,0) -- cycle;
   }
}

\newcommand{\uprightTriangles}{
   \tikz[baseline=(current bounding box.center), scale=0.2]{
       \draw (0, 0) -- (-1, 0) -- (-0.5, -0.86) -- cycle;
       \draw (-0.5, -0.86) -- (0, -1.732) -- (-1, -1.732) -- cycle;
   }
}
\usepackage[ruled]{algorithm2e}
\begin{document}

%\title{Varying exponents  and phase transitions  in the vortex lattice phases of Kagome Ice}
\title{Phase transitions and  crtical exponents in the six-vertex model on kagome lattices}

\author{Wei Zhang }
\affiliation{College of Physics and Optoelectronics, Taiyuan University of Technology, Shanxi 030024, China}
\author{Wanzhou Zhang}
\thanks{zhangwanzhou@tyut.edu.cn}
\affiliation{College of Physics and Optoelectronics, Taiyuan University of Technology, Shanxi 030024, China}
\author{Jie Zhang}
\affiliation{College of Physics and Optoelectronics, Taiyuan University of Technology, Shanxi 030024, China}

\author{Chengxiang Ding}\thanks{dingcx@ahut.edu.cn}
\affiliation{School of Science and Engineering of Mathematics and Physics, Anhui University of Technology, Maanshan, Anhui 243002, China}

\author{Youjin Deng}%
\thanks{yjdeng@ustc.edu.cn}
\affiliation{
Hefei National Laboratory for Physical Sciences at the Microscale and Department of Modern Physics, University of Science and Technology of China, Hefei 230026, China}
\affiliation{Hefei National Laboratory, University of Science and Technology of China, Hefei 230088, China}

\date{\today}

\begin{abstract}Inspired by the experimental realization of direct kagome spin ice  [Yue et al., Nat. Nanotechnol. 19, 1101
(2024)], the theoretical six-vertex model on the kagome lattice is systematically simulated using the directed loop Monte Carlo method. Four distinct vortex lattice phases are identified: (i) antiferromagnetic  leg states and vortex lattice order on both triangular and honeycomb faces, with a winding number $k=1$.  (ii) ferromagnetic  leg states and vortex lattice order on both types of faces, with $k=-2$ on the honeycomb faces and $k=1$ on the triangular faces.  (iii) paramagnetic  leg states  and vortex lattice order  on the triangular faces with $k=1$; and  (iv) paramagnetic  leg states and vortex lattice order on the honeycomb faces with $k=1$. As for ferromagnetic to different types of paramagnetic phase, besides the Ising universality class with $y_t=1$, varying critical exponents have also been found with different values of vertex weights.
The transition between the third  type and fourth  type of vortex lattice phases occurs  with the new exponent  $y_t=1.340(3)$. The third  and fourth types of the vortex lattice phase to the vortex disorder phase are found to be of the Berezinskii-Kosterlitz-Thouless type. These findings contribute to the search for and understanding of ice on complex lattices.

\end{abstract}

\maketitle
%-----------------一，介绍----------------
\section{Introduction}
% 左对齐环境
\begin{sloppypar}

Ice is a substance commonly found in nature and is a fascinating subject for studying various types of phase transitions. There are several distinct forms of ice, such as  water ice~\cite{Salzmann2021}, spin ice observed in natural materials~\cite{Steven}, artificial spin ice~\cite{arm, RevModPhys.85.1473}, and particle-based ice~\cite{cj7}. A key characteristic shared by all these types of ice is the ``ice rule", i.e., the two-in (close) two-out (far away)  constraint~\cite{LinusCarlPauling}.

The Water ice exhibits 19  stable geometric structures, currently identified through high-pressure and low-temperature experiments~\cite{Salzmann2021}.  There are also bilayer water ice experimentally grown on a  surface of Au(111)~\cite{dblayer_ice,Hong2024}. Additionally, artificial spin ices~~\cite{PhysRevLett.129.057202} are also created,  typically using magnetically interacting nanoislands nanowire links\cite{Wang2006}, superconducting-qubit arrays~\cite{AndrewD} and so on. The colloidal particle-based ice \cite{cj1}, is another artificial system capable of making various ices. The colloidal particle ices employ elongated optical traps to confine the colloidal particles. Each optical trap features two wells where the colloidal particles can reside. These arrays of traps can  be  organized into different kinds of lattice~\cite{cj1, cj2, cj3,cj4, cj7, cj8}. Experimentally,  the ice model can also be realized by confining active colloidal fluids in microchannel networks~\cite{jorge2025active, Jorge2024}.

\begin{figure}[t]
\includegraphics[width = 1    \linewidth]{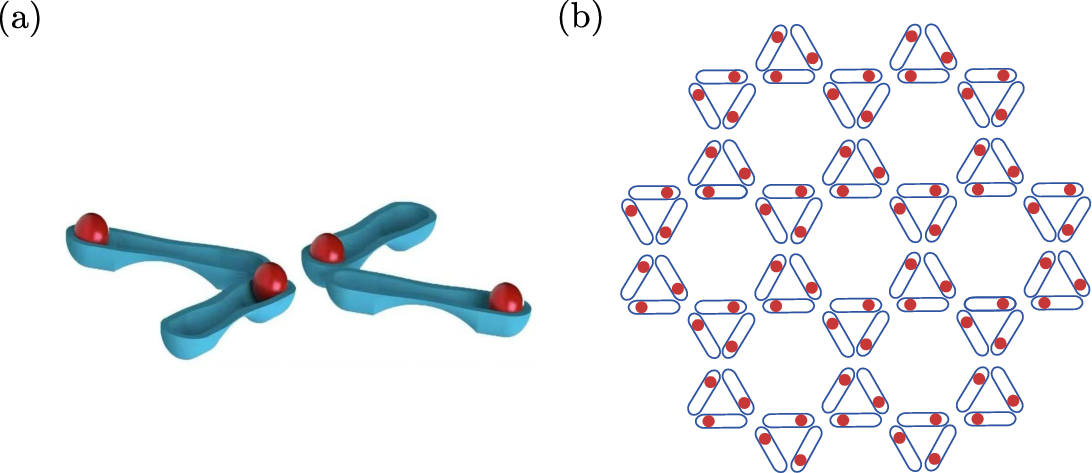}
\caption{The colloidal particle-based ice (a) employs elongated optical traps to confine colloidal particles (b) in an ASI configuration on a Kagome lattice.}

\label{fig:trap}
    \end{figure}

   In ice  naturally formed in nature or created artificially,  Kagome lattice ice, due to its geometric frustration \cite{PhysRevLett.97.257205}, has become a popular lattice for studying artificial ice. The traditional Kagome artificial spin ice (ASI) was constructed by placing nanomagnets on the edges of a honeycomb lattice. In contrast, in 2024, direct experimental realization of Kagome artificial ice was achieved \cite{nature_yw}, where the nanomagnets are placed on the edges of a Kagome lattice. The configurations of the nanomagets can be described by the well-known 16 vertex model. However, the physics of the six-vertex (6V) model remains unclear from the perspective of numerical simulations.

Using the 6V model, the possible vortex lattice (VL) phases~\cite{PhysRevLett.111.067001} and their transition are interesting topics. In certain geometric lattices, such as the bilayer honeycomb lattice~\cite{ dblayer_ice, Hong2024}, the melting of the VL into a vortex disorder phase occurs via a first-order transition. Similar behavior has been observed in real materials such as YBCO and $\text{YBa}_2\text{Cu}_3\text{O}_7$~\cite{firstordervortex, firstordervortex2, Zeldov1995ThermodynamicOO, PhysRevLett.80.4297}. However,  in the simple square lattice, the 6V model exhibits a BKT transition~\cite{Fmodel}. This can be understood. From duality arguments and mapping to Coulomb gas systems, the 6V model, with special parameters, is known to be asymptotically equivalent to the two-dimensional XY model at criticality~\cite{Fmodel}. In the XY model, the unbinding of vortex-antivortex pairs is considered as the cause of the BKT phase transition~\cite{xy1,xy2}.

\end{sloppypar}

\begin{figure}[thbp]  \centering  \includegraphics[width = 1\linewidth]{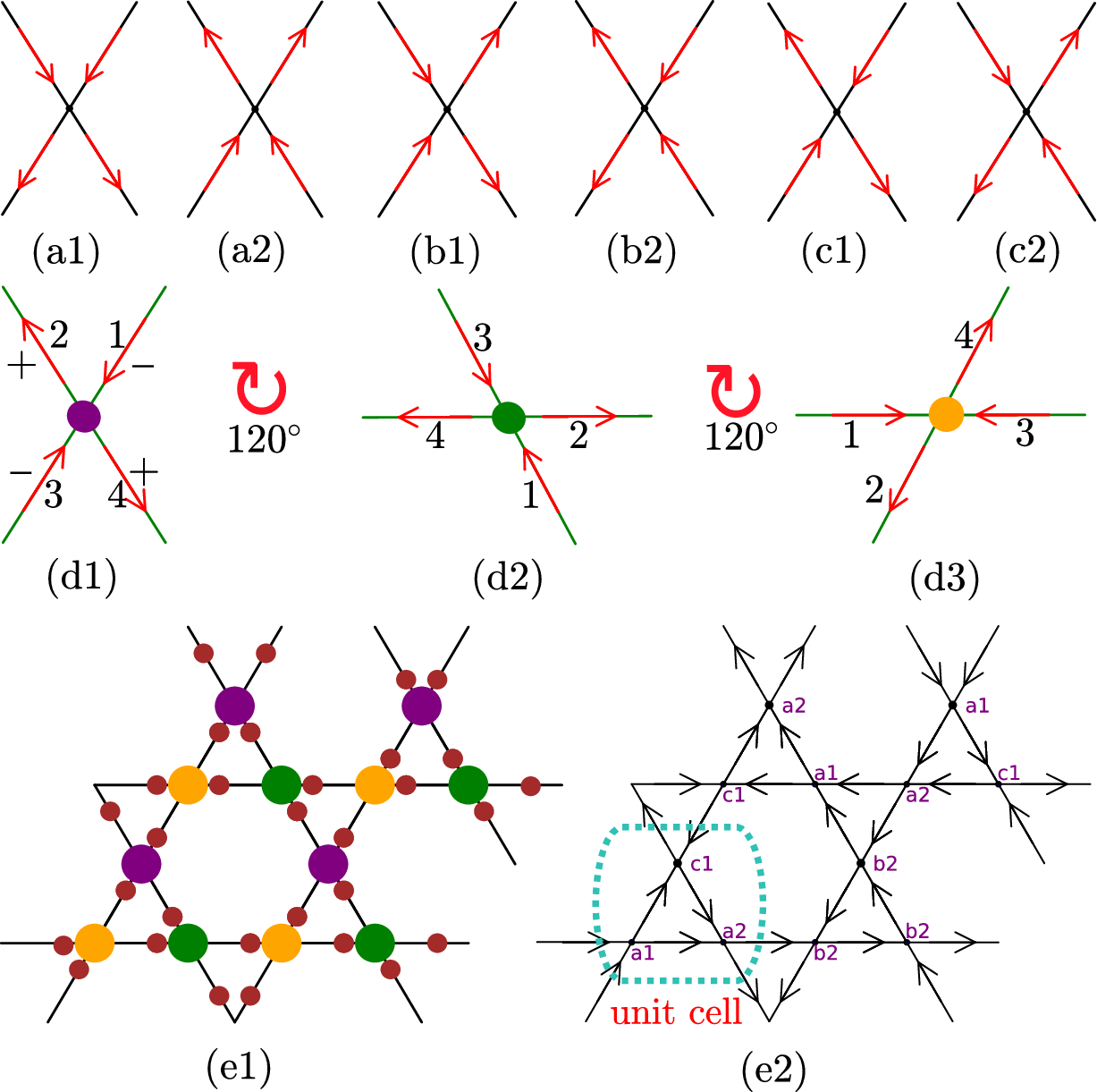}   \caption{Configuration of 6V model. (a1)-(c2) show six types of vertices and their leg configurations; (d1)-(d3) show three types of vertex positions and their leg labels; (e1) shows a possible water ice configuration;(e2) shows $2\times2$ vertex configurations. The unit cells have three vertices and an upright  triangular shape $\mathlarger{\triangle}$.}   \label{fig:6vertex}\end{figure}
%\textcolor{blue}{
A question emerges: {\it Are there interesting vortex lattice phases in the Kagome lattice for the ice model, and what types of transition occur between them?}

To explore this question, we systematically simulate the 6V model on the Kagome lattice using the directed-loop Monte Carlo method~\cite{22in,loop2,loop4,loop5}. The overall phase diagram is obtained from the order parameters related with magnetization and vortices. In the phase diagram, the antiferromagnetic (AFM) and ferromagnetic (FM) phases both coincide with the VL phase, which we further distinguish as VL-I and VL-II. In the paramagnetic(PM)  phase, we observe two other VL phases. Because the vortices exist on triangular and honeycomb faces, the two phases are defined as the VL-T and VL-H phases, respectively, where T and H are the initials of ``triangle" and ``honeycomb".

The details of the phase transitions are also studied carefully. Two distinct critical exponents are obtained for the transition from FM phase to two types of paramagnetic phase: the Ising universality class $y_t=1$   and 
varying critical exponents have also been found for different vertex weight values.
Furthermore, two of the vortex lattice phases  (VL-T and VL-H) melt into the vortex disorder phase via the  BKT-type of transition. The transition between the VL-H and VL-T phases occurs with the exponent  $y_t=1.340(3)$. Experimentally, it could be realized by colloidal particles and trap arrays in a Kagome lattice shape, as shown in Fig.~\ref{fig:trap}.  

The outline of this work is as follows. Sec.~\ref{sec:model} introduces the 6V model on the Kagome lattice, the directed loop algorithm, and the main measured quantities.  Sec.~\ref{sec:mag} shows  the global phase diagram by the magnetization. New exponents are found from the FM phase to the PM phase. Sec.~\ref{sec:vortex}  describes the global phase diagram  by vortex. Conclusive comments and outlook are made in Sec. \ref{sec:con}. %In the appendix~\ref{sec:a} and ~\ref{sec:b}, we confirm our code by the enumeration method and the worm method.

\section{Models, and observed quantities}
\label{sec:model}
%第一部分：模型

\subsection{The short review of the 6V Model} Since the ice rule was introduced in 1935~\cite{LinusCarlPauling}, the 6V model has been extensively studied. In 1967, Lieb~\cite{lieb} calculated the free energy per site exactly. The 6V model is exactly solvable. Over time, the 6 V model has seen several generalizations, such as the 8V, 16V, 19V, and 32V models~\cite{Moore_1983}. It can be mapped to other statistical models, including the Ising model~\cite{six_2_ising}, the Potts model~\cite{potts}, and the Ashkin-Teller model~\cite{six_to_at}. The 6V model can also be mapped to the quantum XXZ chain by taking an anisotropic limit~\cite{XXZ}. For a comprehensive theoretical review of earlier works, refer to the book by Baxter~\cite{baxter1982} (1982) and a detailed review of the integrability of the 6V model~\cite{reviewsix}.

Regarding the 6V model on the Kagome lattice, Baxter (1978) demonstrated that the triangular Potts model is equivalent to the restricted Kagome 6V model~\cite{Baxter1978}. B. Nienhuis (1982) established that the restricted Kagome 6V model is equivalent to the O($n$) model on the honeycomb lattice, and to the Potts model in the triangular lattice~\cite{rotate}.

\subsection{The 6V Model}

Figures \ref{fig:6vertex} (a1)–(c2) present six typical configurations of the 6V model on the Kagome lattice, which satisfy the condition of two incoming and two outgoing. 
 Based on the states of the legs, the type of each vertex can be determined and is represented by the symbols a1-c2.
Compared with the vertices in the shape of a ``$\Huge+$" on the square lattice, the legs pointing upward, downward, to the left, and to the right on the square lattice correspond to the legs numbered 1, 3, 2, and 4 of the Kagome lattice, as shown in Figs.~\ref{fig:6vertex} (d1)-(d3).

The six vertices have their own weights $W_i$ as $a1$, $a2$, $b1$, $b2$, $c1$, $c2$, respectively. In this work, for simplicity, we set $a1=a2$, $b1=b2$, and $c1=c2$.  According to the the Boltzmann distribution, the vertices also have their energy defined as \begin{equation}    \varepsilon_{i} =-\frac{\ln_{}{W {_{i}  }  } }{\beta },  ~~i= 1, \cdots 6\end{equation}where $\beta$ is the inverse temperature, 
 and traditionally, $\beta=1$.

As shown in Fig.~\ref{fig:6vertex} (d1), in mathematical notation, we represent the state of the arrow pointing towards the vertex as a ``-" sign, and the state of the arrow pointing away from the vertex as a ``+" sign.  Considering that in the Kagome lattice, in addition to the upright paired opposite triangles \uprightTriangles, there are also tilted paired opposite triangles \leftTiltedTriangles and  \rightTiltedTriangles. Since the states of the legs at the vertex are closely related to the weight and energy of the vertex, it is necessary to number the legs after strictly tilting them. As shown in Figs.~\ref{fig:6vertex} (d1)-(d3), \leftTiltedTriangles can be obtained by rotating \uprightTriangles clockwise by $120 ^{\circ}$, and \rightTiltedTriangles can be obtained by rotating \leftTiltedTriangles clockwise by another $120 ^{\circ}$. During the rotation process, the numbering of the legs is bounded to the legs themselves. This type of definition was introduced in 1982~\cite{rotate}.

Our theoretical ice model can also inspire first-principles calculations~\cite{PhysRevLett.79.5262} to explore the water ice model on the Kagome lattice. As shown in Figs.~\ref{fig:6vertex} (e1) and (e2), when $H^{-}$ moves away from $ O^{-2}$, an arrow pointing away from the vertex is used to represent it. When $H^{-}$ moves toward $O^{-2}$, an arrow is used pointing towards the vertex. 
Fig~\ref{fig:6vertex} (e2) shows typical vertex configurations with $2\times2$ unit cells, and each unit cell has three vertices.

\subsection{Directed loop algorithm}

In this paper, we apply the directed loop algorithm~\cite{loop2,loop4,loop5}, which has been used  to simulate the sigle-layer~\cite{22in} ice and the bilayer-ice model~\cite{db_ice}. 
A similar loop algorithm is  the worm algorithm~\cite{worm,worm2} with very efficient dynamical behaviours~\cite{d1}.
The 6V model is similar to the flow representation of other models~\cite{flow1,flow2,flow3}, suitable for simulation using the loop types of algorithm.
\begin{figure}[t]%htbp H
   \centering \includegraphics[width = 0.8\linewidth]{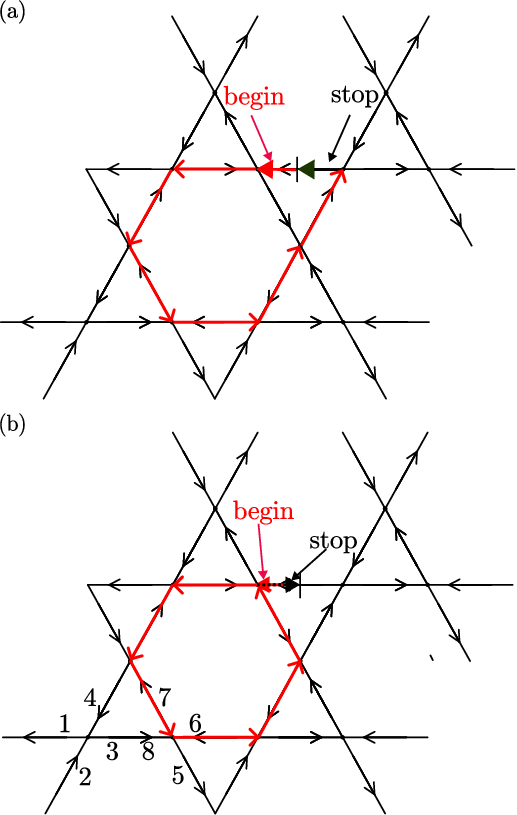}
   \caption{Two types of loop close methods. If one type of loop closure is missing, it will result in longer loops, causing the code to slow down. Similar to the process in quantum Monte Carlo  simulations, a linked list is also constructed. For instance, leg 3 and leg 8 are interlinked with each other.
}
   \label{fig:qyh}
\end{figure}
In the directed loop algorithm, one MC step contains several loop updates. As shown in Fig.~\ref{fig:qyh}, first, randomly select a vertex, and then randomly choose one of the legs, as indicated by the position marked as "begin" or loop head. When the loop head enters the leg corresponding to the vertex, it is called the entrance leg. The leg from which the loop head exits the vertex is called the exit leg. Then, let the loop head move randomly in space, with the probability of the walking direction designed to prevent bouncing. There are two ways for the loop head to close, as shown in Figs.~\ref{fig:qyh} (a) and (b).The loop closes when the exit leg links to the beginning leg. The other way is when the exit leg exactly coincides with the beginning leg.

The probabilities for the various exit legs, given a specific vertex type and an entrance leg, are selected to ensure that the detailed balance condition holds. This results in directed-loop equations. In the solutions, if the bounce probabilities can be avoided, the algorithm's efficiency will be high.

\begin{figure}[htbp]  \centering  \includegraphics[width = 0.8\linewidth]{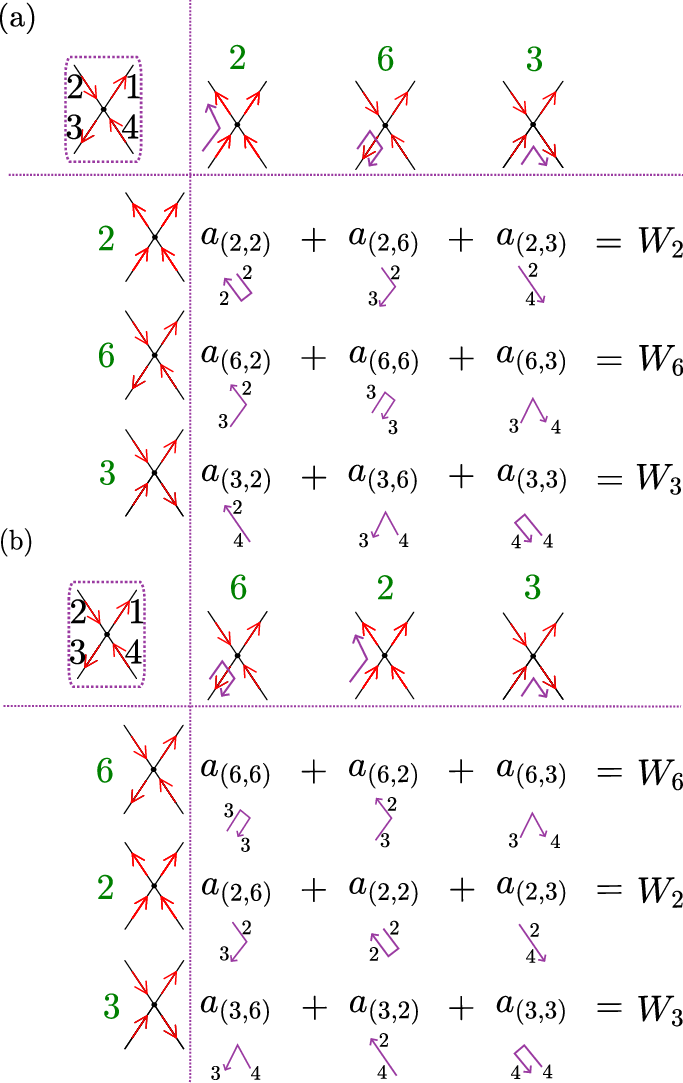}   \caption{   Equation groups of the directed loop probabilities $a_{i,j}$  (a) before sorting of the vertex weight $W_i, i=2, 3, 6$ and (b) after sorting.   \label{fig:loopEq}   }   \label{fig:phase_diagram}\end{figure}

Here, a detailed process of constructing the system of equations is provided for easier follow-up. For convenience, we label the 6V types in Fig \ref{fig:6vertex} (a) as 1 to 6. Suppose we encounter the vertex type 6, and assume that we are entering from leg 3. Then, we attempt to exit from the legs 1, 2, 3, and 4 in sequence. During this process, if leg 1 is used as the exit leg, it results in the formation of a non-existent vertex. As a result, only three valid vertices (2, 6, and 3) can exist. Therefore, the exit legs are 2, 3, and 4 with the entrance leg marked 3,respectively. Using these three vertices, arranged in the order 2, 6, and 3, placed in the top row and the left column as shown in Fig~\ref{fig:loopEq} (a), the transition weights between these vertices correspond to the variables in the 3x3 system of equations. The matrix element $a_{(i,j)}$ represents the weight associated with the transition from type $i$ to type $j$. For the first row, for example, $a_{(2,2)}$, $a_{(2,6)}$, and $a_{(2,3)}$ represent the transition weights from vertex type 2 to vertices 2, 6, and 3, respectively. The corresponding entry leg is fixed at 2, because the transition from vertex 6 to vertex 2 enters via leg 3, but exits via leg 2. Similarly, for the second (third) row, the transitions from vertex 6 (3) to the other vertices have the entry leg fixed at 3 (4). The sum of the elements in each row of $a$ equals the weights of the corresponding reference vertices $W_2$, $W_6$ and $W_3$.

Refs.~\cite{22in,loop2,loop4,loop5} provide the solutions of bounce-free systems of equations. However, the condition requires $W_2>W_6>W_3$, but in reality it is possible that our situation is 
$W_6>W_2>W_3$. 
Therefore, by swapping the first two rows and the first two columns of Fig~\ref{fig:loopEq} (a), a new system of equations can be obtained, as shown in Fig~\ref{fig:loopEq} (b).

To solve for the elements in Fig. \ref{fig:loopEq}(b), for convenience, the positions of the matrix elements are marked with superscripts, for example $a^{(1,1)}=a_{(6,6)}$ and $W^{(1)}=W_6$. 
If $W^{(1)} \le W^{(2)} +  W^{(3)}$, a set of bounce-free solutions is defined as follows:
\begin{equation}
\begin{aligned}
a^{(1,2)}&=\frac{W^{(1)}+W^{(2)}-W^{(3)}   }{2} \\
a^{(1,3)}&=\frac{W^{(1)}-W^{(2)}+W^{(3)}   }{2}\\ 
a^{(2,3)}&=\frac{-W^{(1)}+W^{(2)}+W^{(3)}   }{2} 
\end{aligned}
\end{equation}

If $W^{(1)} > W^{(2)} +  W^{(3)}$, a bounce solution needs to be introduced as
\begin{equation}
\begin{aligned}
a^{(1,1)}&=W^{(1)}-W^{(2)}-W^{(3)}  \\  
a^{(1,2)}&=W^{(2)} \\
a^{(2,3)}&=W^{(3)}   
\end{aligned}
\end{equation}
Then, we normalize these matrix elements, and the normalization formula is as follows:
\begin{equation}
p^{(i,j)}=\frac{a^{(i,j)} }{W^{(i)}} (j=1,2,3)
 \label{eq:oookkk}
\end{equation}

The outgoing leg is determined on the basis of  a random number as follows:
\begin{itemize}   \item If \( 0 \leq \text{rand()} < p_1 \),the outgoing leg is the 1st possible leg that has a chance to exit, though its actual number may be 1 or 2.   \item If \( p_1 \leq \text{rand()} < p_2 \), the outgoing leg is the 2nd possible leg that has a chance to exit   \item If \( p_2 \leq \text{rand()} < 1 \),    the 3rd possible leg that has a chance to exit\end{itemize}The above process determines the outgoing leg in the loop constructing. The algorithm and code are also confirmed by the enumeration method and the worm method in appendices~\ref{sec:a} and \ref{sec:b}.

\subsection{The observable quantities}
%\subsubsection{Magnetic moment}
In order to better characterize the phases and phase transitions of the Kagome lattice,  the following quantities are introduced. 
\subsubsection{specific heat, magnetization, Binder ratio and susceptibility}
\begin{itemize} 
   \item %\textcolor{blue}{
 Specific heat
   \begin{equation}
C_{\text{V}}=\frac{1}{3L^2T^{2}  }  \left (  \left \langle {E}^{2}   \right \rangle -\left \langle {E}  \right \rangle ^{2} \right ) 
\end{equation}
   Here  $T$ denotes temperature, and $\left \langle  \right \rangle $ denotes the ensemble average. The energy of the system ${E}$:
\begin{equation}
{E} =\sum_{i=1}^{3L^2} \varepsilon _{i} 
\end{equation}
where $\varepsilon_{i}$ is the energy corresponding to each vertex and $3L^2$ denotes the total number of vertices and $L^2$ is the number of unit cells.

\item Ferromagnetic magnetization along  the  $x$ direction\begin{equation}M_{\text{F}} =\frac{1}{2L^2}  \left | \sum_{ i=1}^{2L^2} S_{i,x} \right |,\label{eq:mf}\end{equation}where $2L^2$ is the total number of legs along the $x$ direction. $S_{i,x}$ is the state of the legs in the $x$ direction for the vertex $i$. The values of $S_{i,x}$ can be +1 or -1. After testing, the magnetization along the other two directions of the triangle's edges are equivalent to the $x$- direction and do not need to be measured.

\item Stripe magnetization along  the  $x$ direction
\begin{equation}M_{j}^{\text{S}} =\frac{1}{2L}  \left | \sum_{ i=1}^{2L} S_{(i,j),x} \right |,\label{eq:mf}\end{equation}
where, $2L$ denotes the number of stripe legs corresponding to the $j$-th stripe in the $x$-direction, and  $S_{(i, j), x}$  represents the state of the leg at vertex  $i$  in the $ j$ -th stripe along $ x$ direction. $M_{\text{S}}$ is obtained as the average $M_{\text{S}} = \frac{1}{L} \sum\limits_{j=1}^{L} M_{j}^{\text{S}} $.
%这里的||是绝对值的意思&

\item %\textcolor{blue}{ 
Antiferromagnetic  magnetization along  the  $x$ direction
\begin{equation}
M_{\text{A}} =\frac{1}{2L^2}  \left |\sum_{ i=1}^{2L^2} \left ( -1 \right )^{i_x}  S_{i,x} \right |, 
\end{equation}
where $i_x$ is the index of the $i$-th vertex along the horizontal direction, or the column index, in the Kagome lattice.

\item %\textcolor{blue}
Binder ratio corresponding to the magnetization
\begin{equation}
Q=\frac{\left \langle   M^{2} \right \rangle^{2}  }{\left \langle M^{4}\right \rangle } 
\label{eq:q}
\end{equation}
Here $ M$ can be taken as the order parameter $M_{\text{A}}$, $ M_{\text{F}}$ above.

\item %\textcolor{blue}
{Magnetic susceptibility,
\begin{equation}
\chi =\frac{1}{2L^2T^{2}}  \left (  \left \langle M^{2}   \right \rangle -\left \langle M  \right \rangle ^{2} \right )
\end{equation}}
\end{itemize}

%%%%%%%%%%%%%%%%%%%%%%%%%%%%%%%%%%%%%%%%%%%%
\begin{figure}[t]%htbp H
   \centering
  \includegraphics[width = 1\linewidth]{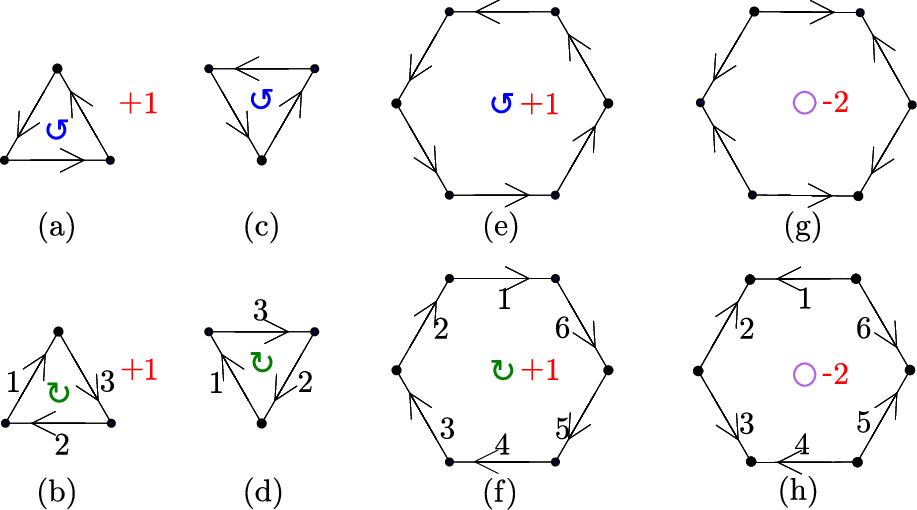}
   \caption{The vortex configurations on the triangular and hexagonal faces are marked with the winding number $k$ in red numbers beside them, and the rotation direction is indicated by colored arrows. (a)-(d) $k=1$, (e)-(f) $k=1$, (g)-(h) $k=-2$.}
   \label{fig:wxdy}
\end{figure}

\subsubsection{vortex and  winding number}
%\textcolor{blue}{
The states of the legs in the Kagome lattice, within the triangular and hexagonal faces, also form interesting VL phases. Similar to the XY model where topological defects are quantified through the vorticity $v$~\cite{PhysRevB.97.045207,Heyang:40503}, the vorticity are also defined in kagome ice:
\begin{equation}
v=\oint_{c}^{} \nabla  \theta \cdot d\vec{l} = 2\pi k,\    k = \pm 1, \pm 2, \dots,
 \label{eq:woxuan}
\end{equation}
where $k$ is the  winding number, its value can be
$ k = \pm 1, \pm 2, \cdots$. The symbol ``c" means the counterclockwise direction on the integral path. 
 %which is  defined  using the direction  each edge of the triangle or honeycomb.
 On a lattice, the integral may be approximated by the sum of the angle differences over a plaquette(triangle or hexagon). To constrain the angular difference within the range $(-\pi,\pi]$,  in the calculation process~\cite{PhysRevB.97.045207,Heyang:40503}, the saw function is used, whose expression is defined as
\begin{equation}
\operatorname{saw}\left(\theta_{i}-\theta_{j}\right)=\left\{\begin{array}{ll}
\theta_{i}-\theta_{j}+2 \pi, & \theta_{i}-\theta_{j} \leq-\pi \\
\theta_{i}-\theta_{j}, & -\pi<\theta_{i}-\theta_{j} \leq \pi \\
\theta_{i}-\theta_{j}-2 \pi. & \pi<\theta_{i}-\theta_{j}
\end{array}\right.
 \label{eq:saw}
\end{equation}
 Through detailed calculations, both the clockwise and counterclockwise vortex winding numbers of the triangle are found to be 1. In Appendix~\ref{sec:c}, an example  is shown to calculate the vortex winding number.

In Figs.~\ref{fig:wxdy}  (a)-(d), the vortex configurations on the triangular faces are shown. The blue represents a counterclockwise vortex and the green represents a clockwise vortex, but both have a winding number $k=1$.
In Figs.~\ref{fig:wxdy}  (e)-(f), the vortex configurations on the honeycomb faces are shown. Both have a winding number $k=1$. However, as shown in Figs.~\ref{fig:wxdy} (g)-(h), both have a winding number $k=-2$.

To further illustrate this, we can consider the angular differences in various geometric configurations to show the characteristics of the vortex. The following three equations describe the sum of the angular differences in the faces of triangular and honeycomb:
\usetikzlibrary{arrows.meta}

% 自定义箭头样式：黑色箭身 + 红色箭头头部
\tikzset{
    myarrow/.style = {
        black,
        -{Stealth[scale=0.8, blue]}, % 仅头部红色
        line width=0.4pt
    },
    myreverse/.style = {
        black,
        {Stealth[scale=0.8, blue]}-, % 反向箭头（仅头部红色）
        line width=0.4pt
    }
}

\begin{subequations}
\begin{align}
% 三角形公式 (a)
\mathlarger{\triangle}~:\sum_{i=1}^{3} \text{saw}(\theta_{\text{mod}(i,3)+1} - \theta_i) &= 2\pi \tag{14a}, \\
% 六边形顺时针公式 (b)
\begin{tikzpicture}[baseline=-0.5ex, scale=0.3]
    \draw[myarrow] (0:1) -- (60:1);
    \draw[myarrow] (60:1) -- (120:1);
    \draw[myarrow] (120:1) -- (180:1);
    \draw[myarrow] (180:1) -- (240:1);
    \draw[myarrow] (240:1) -- (300:1);
    \draw[myarrow] (300:1) -- (0:1);
\end{tikzpicture}
: \sum_{i=1}^{6} \text{saw}(\theta_{\text{mod}(i,6)+1} - \theta_i) &= 2\pi \tag{14b}, \\
% 六边形逆时针公式 (c)
\begin{tikzpicture}[baseline=-0.5ex, scale=0.3]
    \draw[myreverse] (0:1) -- (300:1);
    \draw[myreverse] (240:1) -- (300:1);
    \draw[myreverse] (240:1) -- (180:1);
    \draw[myreverse] (120:1) -- (180:1);
    \draw[myreverse] (120:1) -- (60:1);
    \draw[myreverse] (0:1) -- (60:1);
\end{tikzpicture}
: \sum_{i=1}^{6} \text{saw}(\theta_{\text{mod}(i,6)+1} - \theta_i) &= -4\pi. \tag{14c}
\end{align}
\end{subequations}

\section{Results}
\begin{figure}[hbt]   \centering\includegraphics[width = 0.8\linewidth]{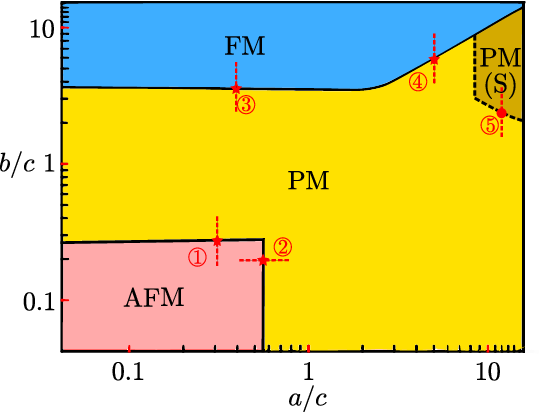}\caption{The schematic representation of the different magnetic phases, including the FM, AFM, PM, and PM(S) region. Due to the presence of regions with striped magnetization  in the PM phase, the brown part is labeled as PM (S). The lines serve as a guide for the eyes, and the points are obtained from finite size scaling. }
% For convenience, the labels \textcolor{blue}{Fig. 8(a)-Fig. 8(c)}are used to mark the positions of the parameters taken for the snapshots drawn later in the phase diagram.}
\label{fig:phase_by_mag}\end{figure}
This model can obtain a phase diagram through the magnetization and another phase diagram through vortices. To simplify, we will discuss them separately.
\subsection{Global phase diagram  by magnetization}
\label{sec:mag}
\subsubsection{Phase diagram and typical  snapshots}
\begin{figure}[htb]
   \centering
\includegraphics[width = 0.7\linewidth]{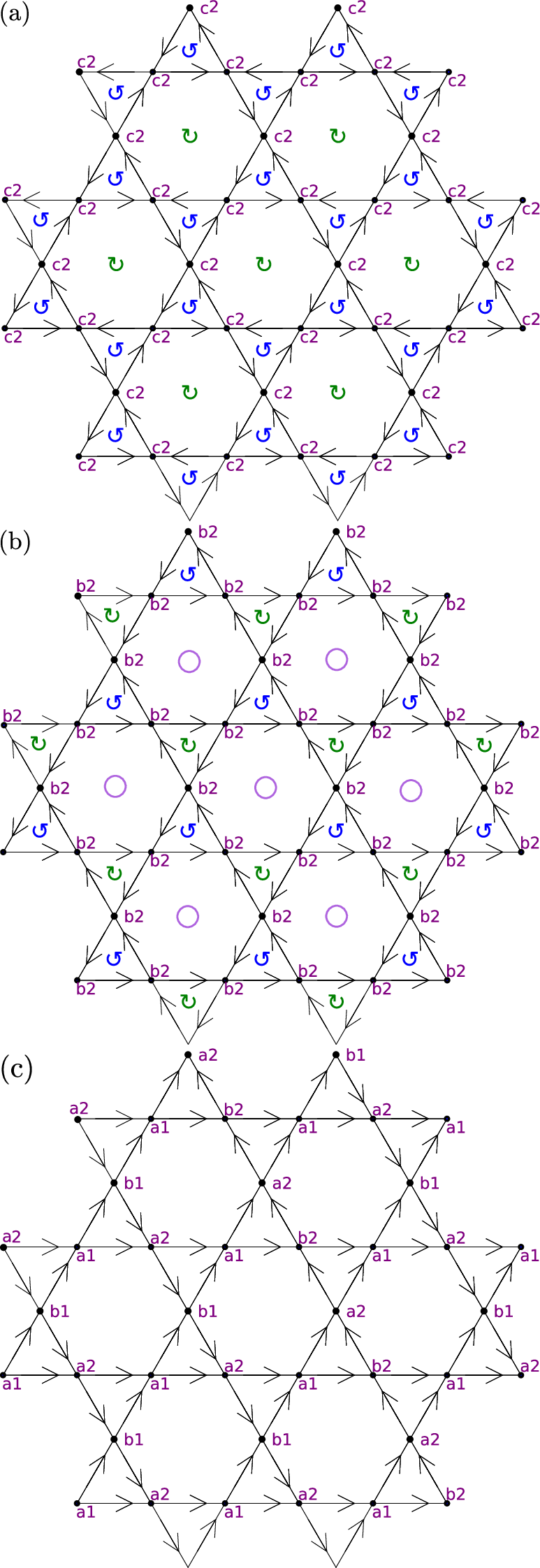}
\caption{Snapshots of (a) AFM phase with winding number $k_\text{T}=1$  and $k_\text{H}=1$. (b) FM phase with $k_\text{T}=1$ and $k_\text{H}=-2$ (c) Striped (PM(S)) region with $k_\text{T}=k_\text{H}=0$. }
   \label{Fig:snapshot}
\end{figure}
% The circles inside the honeycomb represent the vortices formed by the arrows on the six edges, with a winding number of -2.

Figure~\ref{fig:phase_by_mag} shows the global phase diagram, which contains the AFM, FM and PM phase, which includes a stripe region and a conventional region. The phase diagram is plotted in the $a/c-b/c$ plane. In the actual simulation process, we define the weight of the $c$-type vertex as a constant 1, and adjust the ratio of $\frac{a}{c}$ and $\frac{b}{c}$ between 0.05 and 15.

The AF phase occurs in the region where both $a$ and $b$ are smaller than $c$, marked by the light red region,  meaning that in this case the $c2$-type vertices occupy the entire lattice, as shown in Fig.~\ref{Fig:snapshot} (a). In fact, all \( c2 \)-type vertices in the figure can also be replaced by \( c1 \)-type vertices without changing the free energy.

The FM phase occurs in the region where $b$ are larger than $a$ and $c$, marked by the blue region, which means that in this case, the $b2$-type vertices occupy the entire lattice, as shown in Fig.~\ref{Fig:snapshot} (b).
Similarly, all \( b2 \)-type vertices in the figure can also be replaced by \( b1 \)-type vertices without changing the free energy.

 \begin{figure}[t]
    \centering
\includegraphics[width = 1\linewidth]{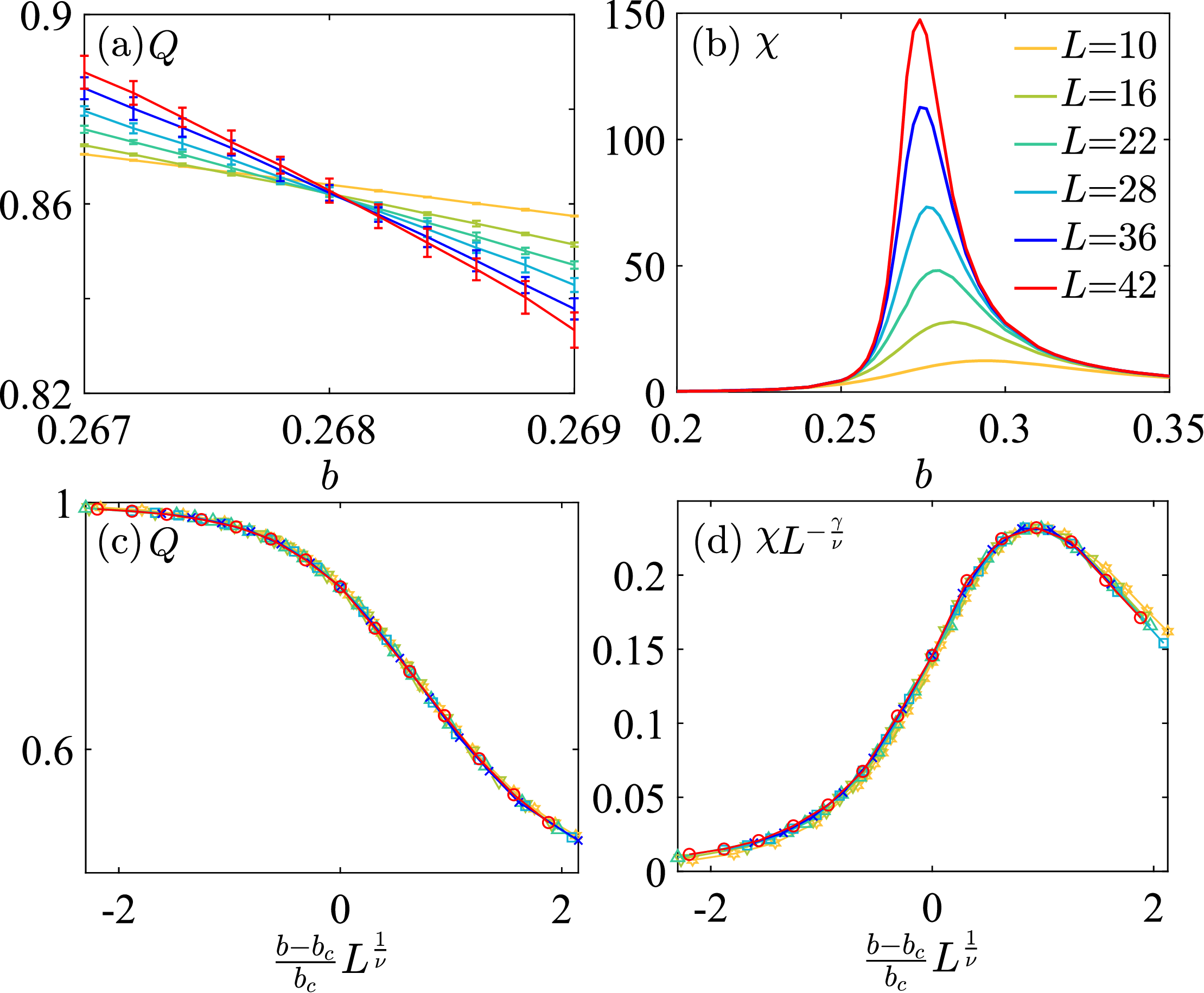}
%\vskip -1 cm相变过程反铁磁到无序态
    \caption{The details along the path  $\textcircled{1}$   (a) $Q$ versus $b$  (b) $\chi$ (c) data collapse of $Q$ (d) data collapse of $\chi$. The exponents $y_t=1$ indicates the phase  transition belongs to the Ising transition.}
    \label{Fig:AFM-D}
\end{figure}

In Fig.~\ref{Fig:snapshot} (c),
the striped region  occurs at the place where both $a$ and $b$ are larger than $c$, as indicated by the brown color.  In the $x$-direction, every leg is in the same state, with each leg pointing either left or right. However, the directions of the arrows between each row are random, and we refer to this as a $x$-striped region, although it still represents a PM  phase.

% \textcolor{blue}{Fig.~\ref{Fig:snapshot} (c) shows a snapshot of a stripe-region configuration, whose emergence is analyzed below:}

% \textcolor{blue}{As evidenced in Fig. ~\ref{Fig:xg1}(b), when the vertex weight of the $c$-type is fixed at unity and the weights of both $a$-type and $b$-type vertices are set substantially larger than one, the presence of $c$-type vertices becomes strongly suppressed across the Kagome lattice, leading to a configuration dominated by $a$-type and $b$-type vertices. }

% \textcolor{blue}{In the Kagome ice system, it is observed that legs 1-3 and 2-4 exhibit a uniform oblique orientation (either upward or downward) in both $a$-type and $b$-type vertices. such directional uniformity stabilizes the formation of a stripe region, thereby suppressing the emergence of other competing orders or region.}

This phase diagram is completely different from that of the square lattice. The phase diagram of the square lattice is symmetric with respect to $a = b$, where $a$ or $b$ being very large can form a FM phase. However, in the Kagome lattice, the symmetry is broken. For example, when $b$ is very large, the b-type vertices can fully occupy the Kagome lattice, forming a FM phase. However, when $a$ is very large, the a-type vertices cannot fully occupy the entire lattice, preventing the formation of a FM phase.

\subsubsection{Ising universality $y_t=1$  at $a=0.3$}
To explore the nature of the AFM-PM phase transition, the physical quantities $Q$ and susceptibility $\chi$ are simulated near the phase transition point, as shown in the path $\textcircled{1}$ in Fig.~\ref{fig:phase_by_mag}.

In Fig.~\ref{Fig:AFM-D} (a), when $a = 0.3$, $Q$ is shown for $L = 10, 16, 22, 28, 36, 42$ as $b$ is scanned along the path. To analyze the data near the critical point $b_c$, finite-size scaling is performed using the following equation:
\begin{equation}\begin{aligned}Q &= Q_0 + e_1(b - b_c)L^{y_t} + e_2(b - b_c)^2 L^{2y_t}  \\ &\quad + f_1 L^{-\omega},
\end{aligned}\label{eq:q}\end{equation}
where $y_t$ is the thermal exponent, and $e_1$, $e_2$, $f_1$ are non-universal expansion coefficients and $\omega$ is fixed to 1.  The fitting using the least- squares method gives $b_c =0.268(1)$. The critical exponent is $y_t=1.01(2)$ indicating the Ising universality. Based on the estimated precise phase transition point and critical exponents, we also perform data collapse, where the horizontal axis represents $tL^{y_t}$ and the vertical axis represents $Q$, where $t$ is the reduced parameter defined as $\frac{b-b_{c}}{ b_{c}}$. The $Q$ \textcolor{blue}{curves} of the different sizes  collapse nicely onto each other, as shown in Fig.~\ref{Fig:AFM-D} (c).

To calculate the critical exponent $\gamma$, the following scaling hypothesis formula is used,
\begin{equation}
\chi_{max} \propto|t|^{-\gamma}
\propto L^{\gamma/\nu}.
\label{eq:xx}
\end{equation}
Taking the logarithm of both sides of the above formula, the corresponding slope is the exponent $\gamma/\nu$, and its fitted value is $1.76(1)$, which is consistent with the Ising exponent $\gamma=7/4=1.75$ within the errobar. %A slight deviation, possibly due to the effect of size.

Using the evaluated exponent \( \gamma \), we also perform data collapse by plotting \( \chi L^{-\gamma/\nu} \) versus \( tL^{1/\nu} \), where the curves for different system sizes overlap very well~\cite{Sandvik2010ComputationalSO}. Using the following scaling relations,
\begin{subequations}
\begin{align}
\alpha +2\beta +\gamma  =2
 \tag{\theequation a}, 
   \label{aaaaa}                \\  
\gamma =\beta (\delta -1)
 \tag{\theequation b},
  \label{bbbbbbb}\\
 \nu d  =2-\alpha \tag{\theequation c},
 \label{ccccccc}\\
 \gamma =\nu (2-\eta ),\tag{\theequation d}
 \
\end{align}
\label{dddddddd}
\end{subequations}
the remaining exponents \( \alpha \approx 0 \), \( \delta\approx 15 \), \( \eta \approx \frac{1}{4} \), and \( \beta \approx 0.125 \) can be obtained. These critical exponents are consistent with Ising universality.

By  extensive calculations and finite-size scaling, we demonstrate that both path $\textcircled{2}$ (the AFM-to-PM transition) and path $\textcircled{3}$ (the FM-to-PM transition) manifest Ising-type critical behavior. For simplicity, the  results are not presented explicitly herein.

\subsubsection{Critical exponent variation at $a=0.4, 4,5,$ and $6$ for the FM-PM transition}

\begin{figure}[t]
   \centering
\includegraphics[width = 1\linewidth]{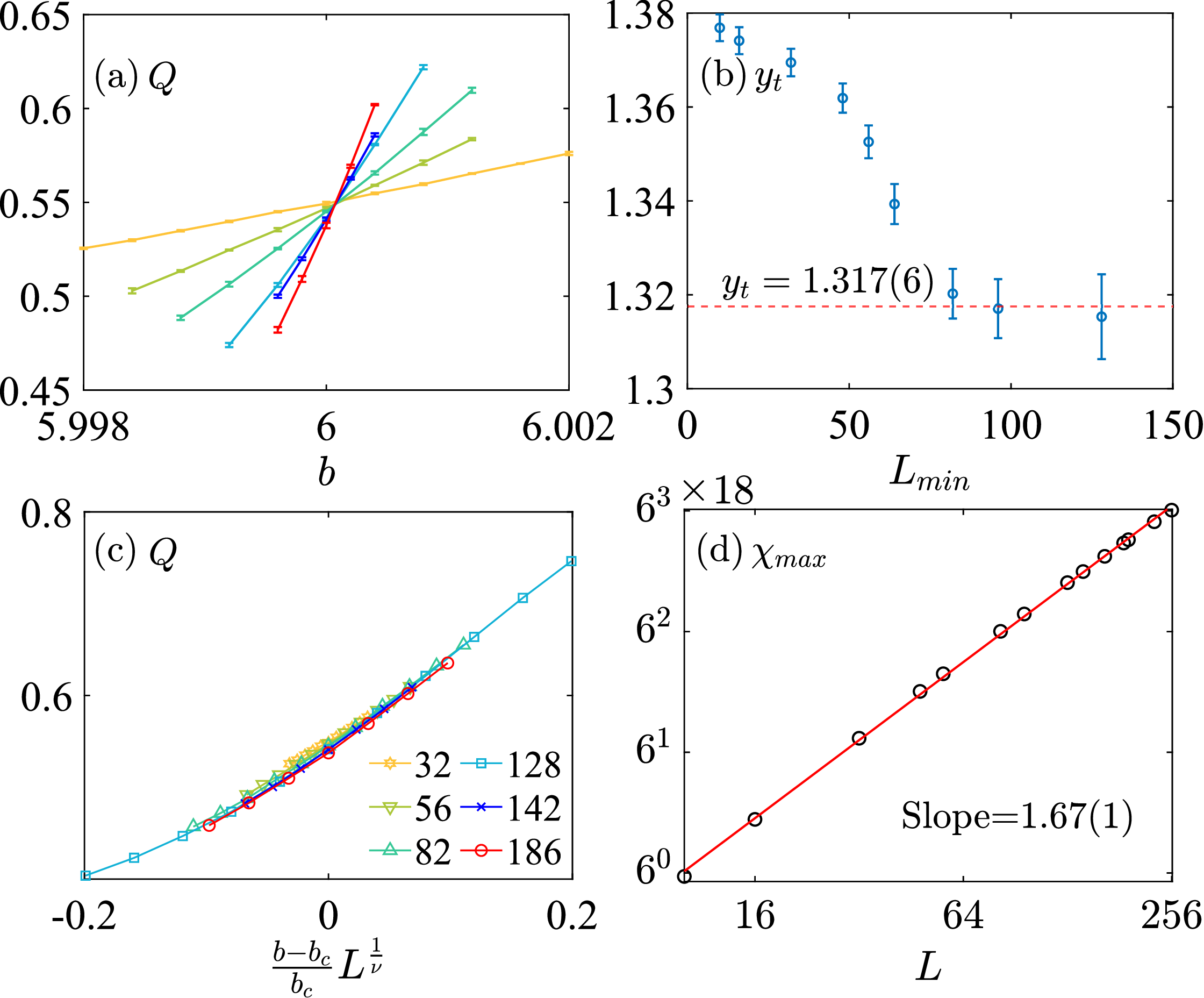}
   \caption{Details along the path $\textcircled{4}$
(a) the data lines of \( Q \) versus \( b \) for different system sizes.
(b) with $L_{max}=256$, the fitted $y_t$ versus $L_{min}$ and become stable at $L_{min}=82$.
Finite-size  analysis of Eq.~\ref{eq:q} yield reasonable values of the critical exponent, \( y_{t} = \frac{1}{\nu} \approx 1.317(6) \).
(c) data collapse of $Q$ and 
(d) $\chi_{max}$ versus $L$ in log-log plot.} 
   \label{Fig:FM-D}
\end{figure}

As the parameter $a$ changes from $a=0.4$ to $a=6$, the critical exponent changes, even though the phase transition is fixed from the FM to PM.

Figure~\ref{Fig:FM-D} (a) shows the lines $Q$ .vs. $b$ in the regimes of $b_c$ in the range $5.998< b < 6.002$, with various system sizes from $L=10-186$. The data for \( L = 256 \) have also been calculated, but it is not presented here as it closely resembles the data for \( L = 186 \). In Fig~\ref{Fig:FM-D} (b), using the data $Q$ .vs. $b$ and the least squares method as given in Eq.~\ref{eq:q}, the critical weight parameter $b_c$ and the critical exponent $y_t$ are obtained, beginning with different values  of  $L_{\text{min}}= 10, \cdots, 128$ and  the fixed  maximum size  $L_{\text{max}}=256$.

In Table \ref{Tab:Q}, when \( L_{\rm min} \) increases from 10 to 128, the estimated critical point remains stable around \( b_c = 6 \). Furthermore, the residual \( \chi^2 \) per degree of freedom is close to 1, which is reasonably acceptable. Therefore, the final quoted value of \( b_c \) is taken as \( 6.000\,09(9) \).
In addition, taking into account that additional corrections might not be included in the fitting formula, 
we double the statistical fitting error in the final estimate $b_c=6.000\,0(2)$.
The fitting result for \( y_t \) becomes convergent when \( L_{\rm min} \) increases to 82. By averaging the fitted values \( y_t \)  for \( L_{\rm min} = 82, 96, 128 \), and using an error bar that is three times the standard deviation, our final result is \( y_t = 1.317(6) \).
\begin{table*}[hbt]
    \caption{Fitting results for the Binder ratio $Q$ using the ansatz Eq.~\eqref{eq:q}.}
    \tabcolsep=0.2 cm
    \begin{tabular}{cclllllllll}
    \hline
    \hline
    Obs. & $L_{min}$ & $\chi^2$/DF & $b_c$ & $y_t$  & $Q_0$ & $e_1$ & $e_2$ & $f_1$ & $\omega$ \\
    \hline
    $Q$  
    % & 64  & 25.5/34 & 0.109~20(1)  & 0.66(1) & 1.444~2(6) & -2.6(1) & 5(1) & 0 & 0 & 0  \\
    % & 96  & 21.0/22 & 0.109~22(2)  & 0.66(2) & 1.445(1)  & -2.5(3)  & 4(1) & 0 & 0 & 0  \\
    & 10 & 8.0412784152 &6.000~07(1)  & 1.376(2) & 0.5495(1) & 0.115(1) & 0.0136(5) & 0.017(2)  & 1 \\
    & 16  & 3.3905907513 & 6.000~06(1)   & 1.374(2) & 0.5471(1) & 0.116(1) & 0.0152(5) & 0.105(4) & 1\\
  
    & 32 & 3.0189990861  & 6.000~06(1) & 1.369(2) & 0.5469(3) & 0.119(1) & 0.0159(5) & 0.11(1)  & 1  \\
    & 48  & 2.6083228630 & 6.000~06(2)  & 1.361(3)  & 0.5475(4) & 0.123(1) & 0.0169(6) & 0.07(2) & 1 \\
    & 56  & 2.4206931032& 6.000~06(2)  & 1.352(3)  & 0.5478(5) & 0.129(2)  & 0.0184(7) & 0.06(3) & 1  \\

     & 64  & 2.1953299937 & 6.000~07(3)  & 1.339(4)  & 0.5484(7) & 0.138(2)  & 0.021(1) & 0.01(4) & 1 \\
      & 82  & 1.6517468039 & 6.000~07(4)  & 1.320(5)  & 0.548(1) & 0.152(3)  & 0.025(1) & 0.01(8) & 1\\
      
      & 96  & 1.7174687795 & 6.000~07(4)  & 1.317(7)  & 0.549(1) & 0.155(4)  & 0.026(1) & -0.1(1) & 1 \\
      
       & 128  & 1.6734930484 & 6.000~09(9)  & 1.315(9)  & 0.556(3) & 0.158(7)  & 0.027(2) & -0.7(3) & 1\\
    \hline
    \hline
    \end{tabular}
    \label{Tab:Q}
\end{table*}

In Fig.~\ref{Fig:FM-D} (c), based on the estimated results about $b_c$ and $y_t$,  data collapse is performed, where the horizontal axis represents $\frac{b-b_{c}}{ b_{c}}L^{1/\nu}$ and the vertical axis represents $Q$, and the data from different lattice sizes overlap very well onto a single curve.
In Fig.~\ref{Fig:FM-D} (d), using the relationship in Eq.~\ref{eq:xx}, $\chi_{\text{max}}$ vs $L$ is plotted on a double logarithmic scale.  The slope is obtained as $\frac{\gamma}{\nu} = 1.67(1)$. Using the estimated $y_t$, $\gamma = 1.26(1)$ is obtained. The error is calculated using the error propagation formula. For the exponent $\beta$,  we use the scaling of the square root of magnetization defined in Eq.~\ref{eq:mf} as follows. \begin{equation}M_{\text{rms}}^{b = b_c}\propto \sqrt{L^{2-d-\eta}} \propto L^{\frac{-\beta}{\nu}},\end{equation}where $d=2$ is the dimension of the space. The detailed derivation is shown in ~\cite{Fss_1}. Using a fit, we get  $\frac{\beta}{\nu} = 0.189(4)$, and finally $\beta = 0.14(2)$ is obtained. We also determine $\alpha = 0.482(8)$ from Eq.~\ref{aaaaa}. Finally, the exponents $\alpha$, $\beta$, $\gamma$, $\eta$ are obtained.

The critical exponents $y_t$ for the transition from the  PM to FM phases at $a$ = 4 and $a$ = 6 are measured to be 1.164(3) and 1.381(1), respectively.
 %The corresponding data and discussions are provided in Appendix ~\ref{sec:d}.}
 \begin{figure}[hbt] % 这里使用figure环境是为了让内容可
\includegraphics[width=0.5\textwidth]{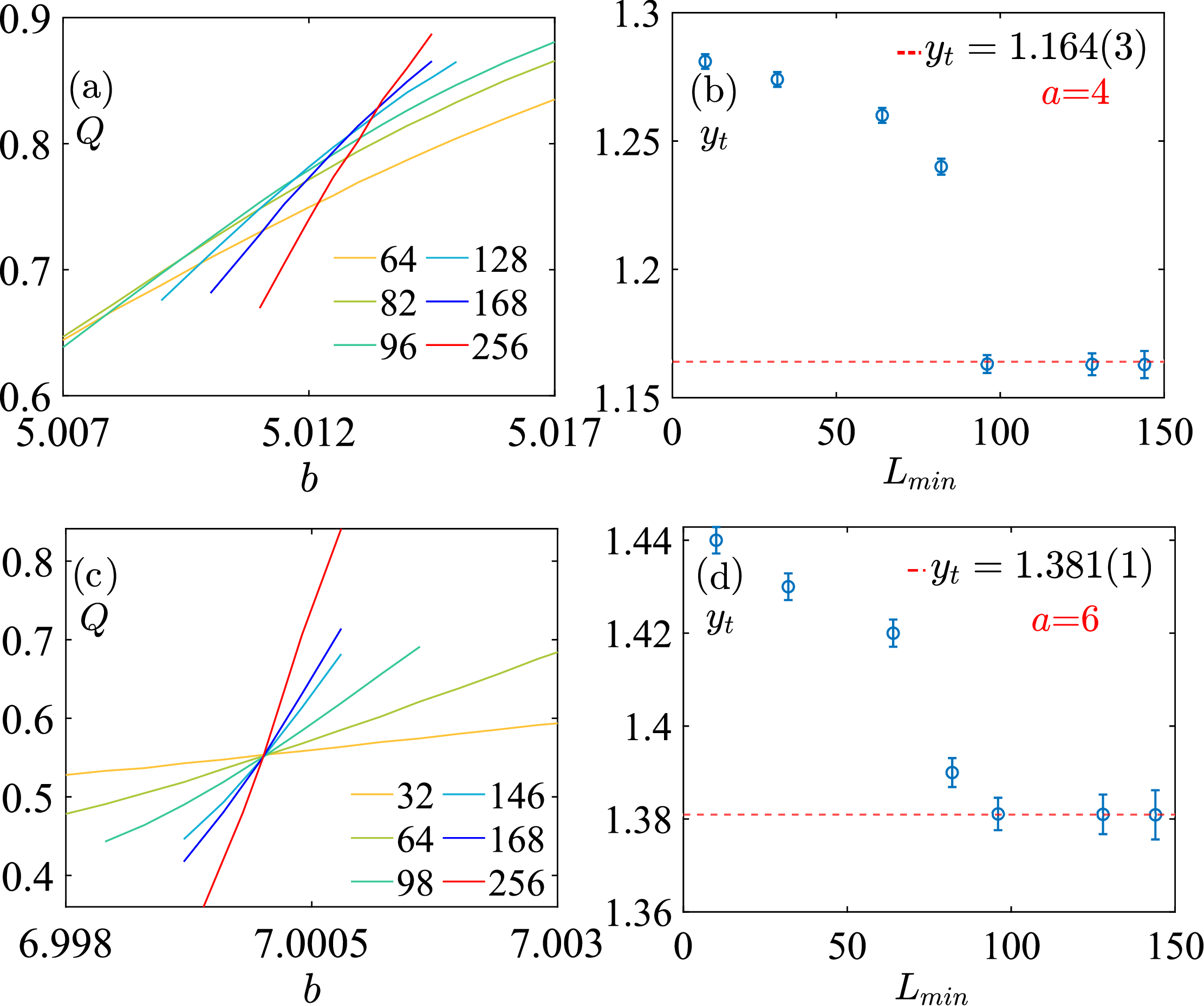}  
    \caption{The Binder ratio $Q$ around the  critical points and fitting details using different $L_{\text{min}}$ (a) the data of $Q$ of fixing $a=4$ and scanning $b$ (b) $y_t$ of using different values of $L_{\text{min}}$ (c) the data of $Q$ of fixing $a=6$ and scanning $b$ (d) $y_t$ of using different values of $L_{\text{min}}$. } \label{fig:bcsj}
  \end{figure}
  In Fig.~\ref{fig:bcsj}, we present the Binder ratio $Q$ and critical exponents $y_t$ obtained by scanning parameter \( b \) for two different values of parameter \( a \), along the FM-PM boundary.  At $a=4$,  the value is $y_t=1.164(3)$ and at $a=6$, the value is $y_t=1.381(1)$.  The trends of these quantities clearly reflect the variation of critical behavior with the parameter \( a \) near the phase transition.
The fitting method is the same as that for the path $\textcircled{4}$. As shown in the fitting results (see Figure \ref{fig:bcsj} (b) and (d)), among the data used for fitting, the minimum system size is 32 or 64, and the maximum system size is 256. When the varied \( L_{\text{min}} \) is less than 96, the critical exponent \( y_t \) does not converge; in contrast, all these critical exponents achieve convergence once \( L_{\text{min}} > 96 \).

In total, for \( a = 4, 5, 6 \), the values of the critical exponent \( y_t \) are determined to be 1.164(3), 1.317(6), and 1.381(1), respectively. This observation naturally raises an open question: Does \( y_t \) converge to a well-defined value as \( a \) increases further? However, a practical challenge emerges here—with the increase of \( a \), achieving loop closure becomes increasingly difficult in large-size Kagome ice simulations that adopt the directed loop algorithm. Given that this question exceeds the  scope of the current work, it is reserved for further exploration and resolution in future investigations.

\subsubsection{Stripe region, a specific form of PM state.}
\begin{figure}[htbp]
    \centering
\includegraphics[width = 1\linewidth]{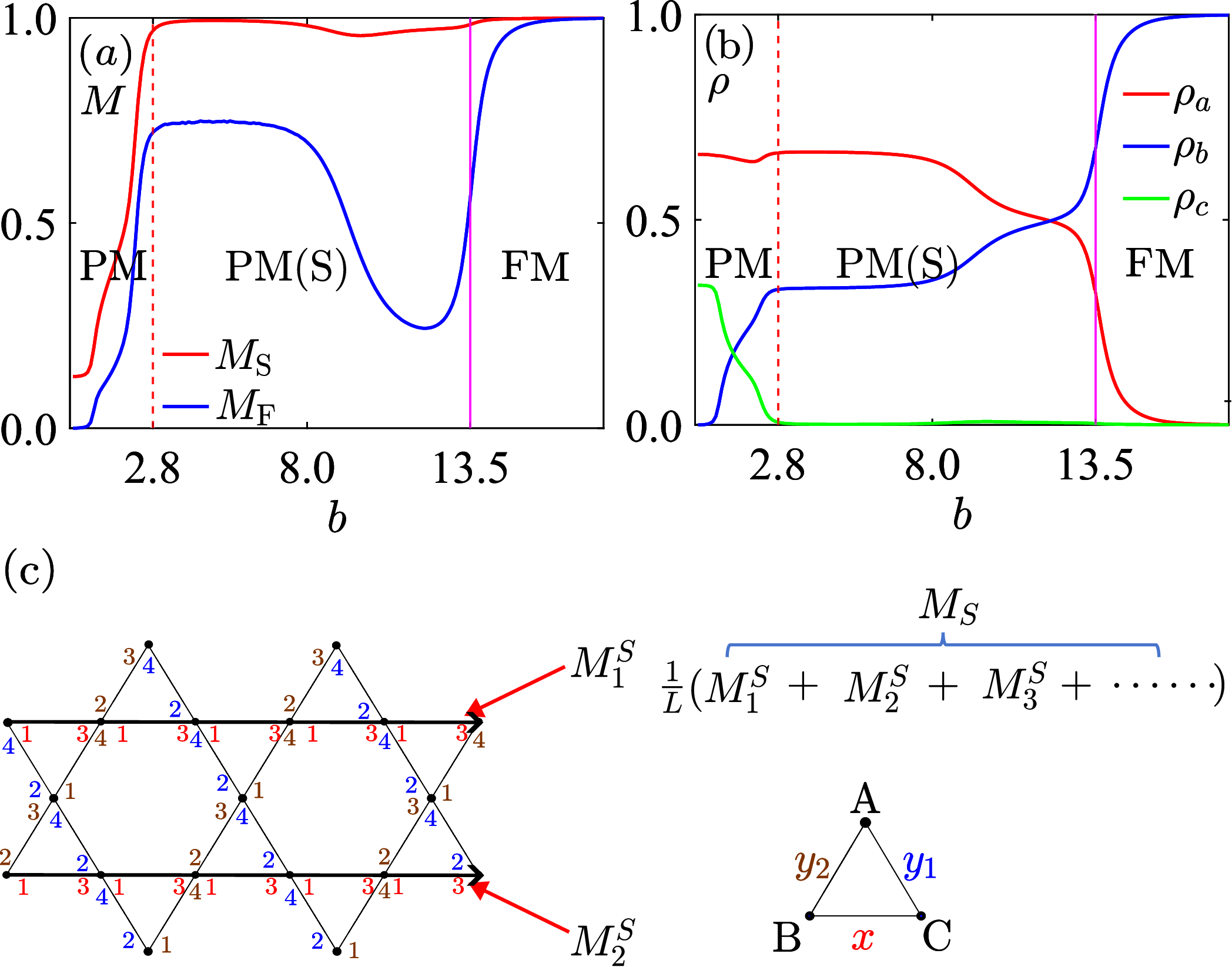}
%\vskip -1 cm相变过程反铁磁到无序态
    \caption{The details of the PM(S) regions and the magnetization as function of  $b$ with  $L$=10 and vertex weight $a$=12. ‌  (a) $M_\text{S}$ and $M_\text{F}$ versus $b$. (b) $\rho_a$, $\rho_b$ and $\rho_c$ versus $b$. (c)  Numbering of vertex legs on the kagome lattice and  schematic diagram of striped magnetization. The stripes can form along he three directions labeled 
 $x$, $y_1$, and $y_2$.}
    \label{Fig:xg1}
\end{figure}

The stripe region belongs to the PM phase, but is different from the pure PM phase. Along each direction, for example, the horizontal direction, all legs are in the same state. However, the direction of each row is random. To quantitatively describe this feature, we have plotted $M_\text{S}$ and $M_\text{F}$, with the parameter being $a = 12$
and $L=10$ as functions of the parameter $b$,  presented in Fig.~\ref{Fig:xg1} (a).
In the range $b<2.8$, it is the pure PM phase with both $M_\text{F}<1$ and $M_\text{S}<1$. However, in the range $2.8<b<13.5$, $M_\text{S}\approx$ 1 while $M_\text{F}<1$.  Since the transition (path \textcircled{5} in Fig.~\ref{fig:phase_by_mag}) between the pure PM region and  stripe region is not a true phase transition, we just used the above quantitative conditions to distinguish them.
In the range $b> 13.5$, $M_\text{F} \approx $ 1 and $M_\text{S} \approx$ 1, it indicates that it is the FM phase and long-range magnetic order.

Now we try to understand the stripe region using  the three types of vertex density, 
\begin{equation}
    \rho_a = \frac{1}{3L^2} \sum_{i=1}^{3L^2} n_i^a,
\end{equation}
 $\rho_a$, $\rho_b$, $\rho_c$, where 
where $3L^2$ is the total number of vertices, and $n_i^a = 1$ if vertex $i$ belongs to a type-$a$ vertex and $n_i^\alpha = 0$ otherwise.
$\rho_b$ and  $\rho_c$ are defined similarly. 
In Fig.~\ref{Fig:xg1}(b), $\rho_a$, $\rho_b$ and $\rho_c$ vary with $b$ are shown.
As  $b$ increases,  $\rho_c$ decreases until $\rho_c \approx$  0 at $b$ = 2.8, at which point the lattice is fully filled by vertices of types $a$ and $b$.  There are no other possibilities such as $aac$ and $bbc$.

To further describe the configuration of $aab$ and its striped properties, the leg statuses along the three directions of the vertices are discussed. The three directions are labeled 
 $x$, $y_1$, and $y_2$ as shown in  triangle $ABC$
in the right in Fig.~\ref{Fig:xg1} (c), where 
the triangle represents the unit cell of the kagome lattice.
In the $x$ direction, the leg labels are 1 and 3, which, as known from Figs.~\ref{fig:6vertex} (a1)--(c2), have the same arrow directions for both $a$-type and $b$-type vertices.
Similarly, in the $y_1$ direction, the leg labels are 2 and 4, which have the same arrow directions for both $a$-type and $b$-type vertices.
 For the  $y_2$ direction, the labels of the legs are  3-1-4-2 , where   3 and 1  legs of vertex \uprightTriangles and  4 and 2 legs of
\rightTiltedTriangles are collected together. The arrow directions for the legs 3-1-4-2 are the same again.

% \begin{subequations}
% \begin{align}
% \rho_{a} &=\frac{1}{N} \sum_{i=1}^{N_a}n_{(i,a)}, 
%    \label{roua}                \\  
% \rho_{b} &=\frac{1}{N} \sum_{i=1}^{N_b}n_{(i,b)},
%   \label{roub}                  \\ 
%   \rho_{c} &=\frac{1}{N} \sum_{i=1}^{N_c}n_{(i,c)},
%   \label{rouc}
% \end{align}
% \label{djmd}
% \end{subequations}

\subsection{Phase diagram  by vortex}
\label{sec:vortex}
\subsubsection{vortex lattice phases}

\begin{figure}[htbp]  \centering  \includegraphics[width = 0.8\linewidth]{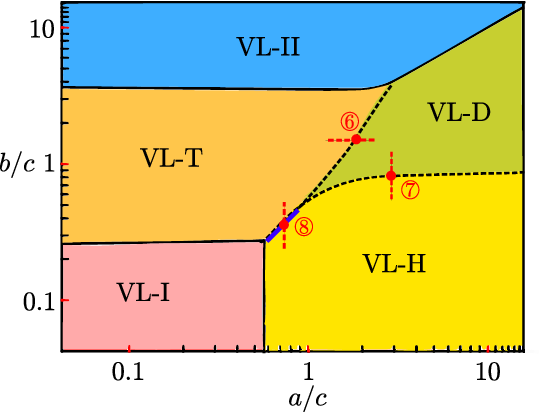}   \caption{Phase diagram and four types of perfect vortex lattices. The lines are guides to the eye, and the points are obtained through finite-size scaling.}
% For convenience, the labels (a)-(e) are used to mark the positions of the parameters taken for the snapshots drawn later in the phase diagram.\textcolor{red}{For convenience, the labels Fig. 8(a)-Fig. 13(b) are used to mark the positions of the parameters taken for the snapshots drawn later in the phase diagram}}
\label{fig:phase_figure_wx}\end{figure}

\begin{figure}[ht]
   \centering
\includegraphics[width = 0.7\linewidth]{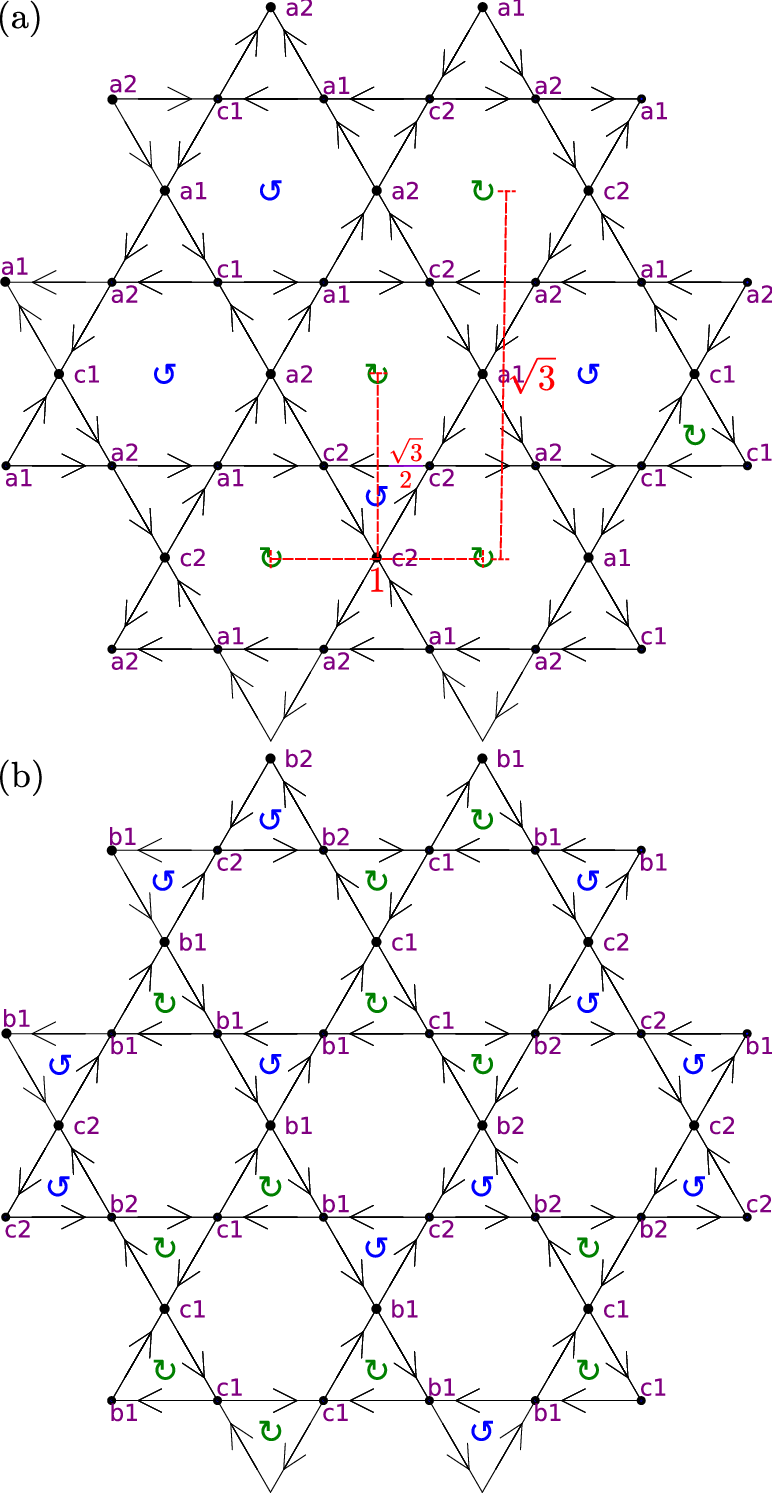}
\caption{Snapshots of the  (a) VL-H and (b) VL-T  phases.
}
\label{fig:snapshotab}
\end{figure}

In Fig.~\ref{fig:phase_figure_wx},  the phase diagram  is shown by vortex-related quantities. The phase diagram includes various VL phases. The VL phases are classified as the VL-I phase, VL-II phase, VL-T phase, VL-H phase, and VL-D phase.

For the first type of VL phase, marked by VL-I phase, the snapshot is shown in Fig.~\ref{Fig:snapshot} (a). Regardless of whether the triangles are upright or inverted, the vortices rotate counterclockwise. In hexagonal faces, the vortices rotate clockwise. The winding number is $k=1$ in both cases, indicating a perfect vortex lattice phase. This phase has been found experimentally in Ref.~\cite{nature_yw}, and called ``Ferrotoroidicity".

For the second type of VL phases, marked by the 
VL-II phase, the snapshot is shown in Fig.~\ref{Fig:snapshot} (b).
Adjacent triangular faces have positive vortices with a winding number of 1, but they rotate in opposite directions. In contrast, in the hexagonal unit cell, the vortices are antivortices with a negative winding number of -2. 
The circles inside the honeycomb face represent the vortices formed by the arrows on the six edges, with a winding number of -2.
Although the same vortex configuration has been observed in Ref.~\cite{song2}, we are the first to discover this phase in the 6V model.

For the third type of VL phase, marked by the VL-H phase, the snapshot is shown in Fig.~\ref{fig:snapshotab} (a). The VL phase is a partial VL order, as  vortices are present only in the honeycomb cells, while the triangular cells contain no vortices. In each unit-cell,  the ratio of the number of $a$-type vertices to that of $c$-type vertices is 2:1. 

For the fourth type of VL phase, marked by the VL-T phase, the snapshot is shown in Fig.~\ref{fig:snapshotab} (b). The VL-T phase is also a partial VL order, as vortices are present only in the triangular faces, while the honeycomb faces contain no vortices. This configuration also exists in the frustrated Kagome XY antiferromagnet~\cite{song1}.

\begin{figure}[t]%htbp H
   \centering
    \includegraphics[width = 1\linewidth]{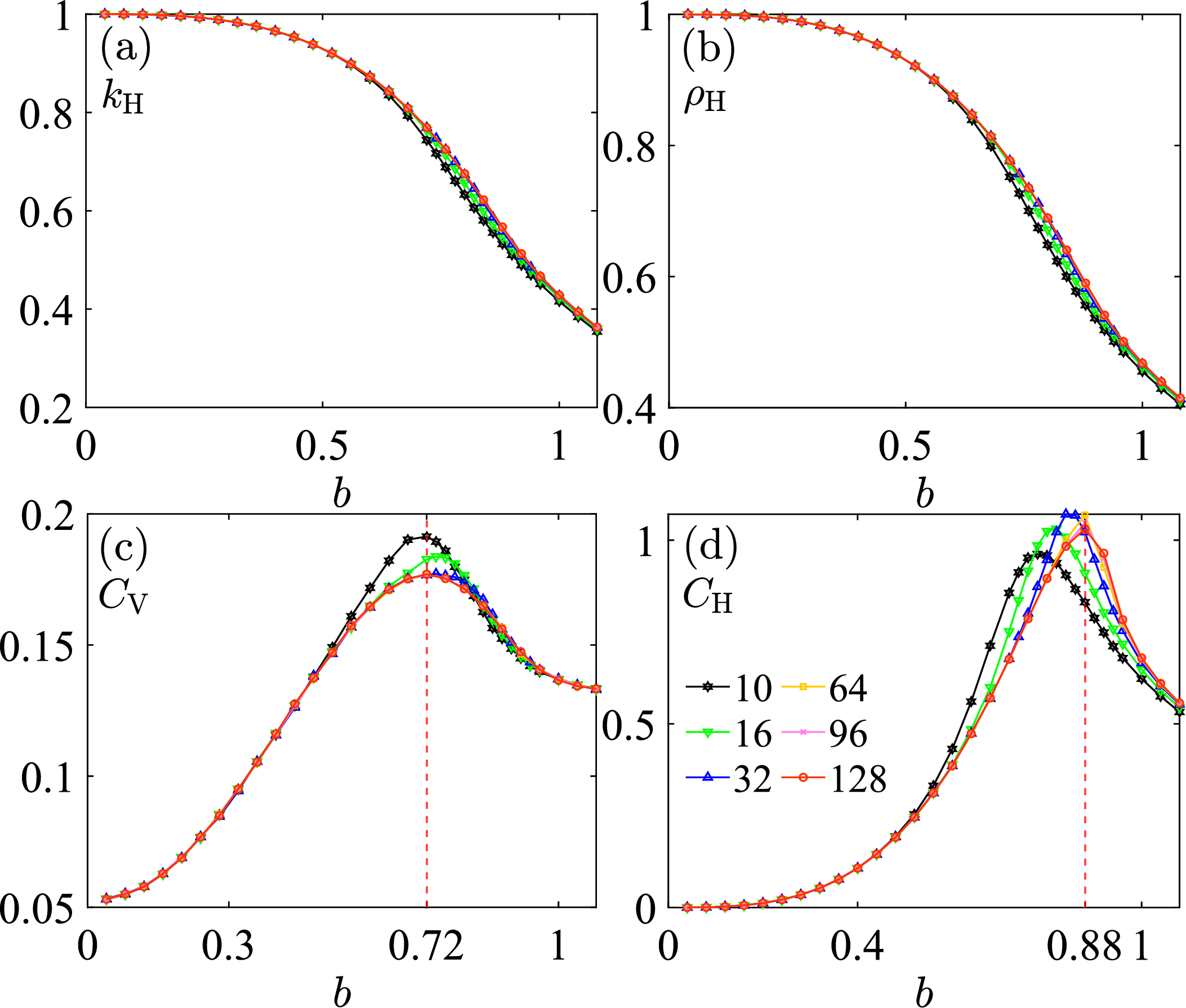}
   %图（a）和图（c）分别是由相（c）到相(f)测量的涡旋1的涡旋密度和涡旋密度的比热
  \caption{ 
Details of the quantities along the path \textcircled{7}: (a) the winding number density \(k_\text{H}\), (b) vortex density \(\rho_\text{H}\), (c) specific heat \(C_\text{V}\), and (d) a specific heat-like quantity \(C_\text{H}\). The convergence of the specific heat indicates the BKT transition.}
   \label{fig:14}
\end{figure}

\subsubsection{BKT transitions}

Due to the presence of vortices in the system, it is necessary to explore whether a BKT transition exists. Therefore, to detect such transition, it is necessary to define the vortex density as follows

\begin{subequations}
\begin{align}
\rho_{\text{H}} &=\frac{1}{L^2} \sum_{i=1}^{L^2}n_{i}, 
   \label{rouH}                \\  
\rho_{\text{T}} &=\frac{1}{2L^2} \sum_{i=1}^{2L^2}n_{i},
  \label{rouH}
\end{align}
\label{dddddddd}
\end{subequations}
\noindent where \( L^2 \)($2L^2$) denotes the total number of honeycomb(triangles)  faces, \( n_i = 1 \) when vortices are present, and \( n_i = 0 \) if they are absent. The definition of the vortex density \( \rho_\text{T} \) for the triangular  faces is similar. We also define quantities similar to the specific heat in order to determine the transition point of the BKT transition, as following,
\begin{subequations}
\begin{align}
{C}_{\text{H}}  &=\frac{L^2}{T^{2}  }  \left (  \left \langle \rho_{\text{H}}^{2}   \right \rangle -\left \langle \rho_{\text{H}} \right \rangle ^{2} \right ), 
   \label{CvH}                \\  
   {C}_{\text{T}}  &=\frac{2L^2}{T^{2}  }  \left (  \left \langle \rho_{\text{T}}^{2}   \right \rangle -\left \langle \rho_{\text{T}} \right \rangle ^{2} \right ).
  \label{CT}
\end{align}
\label{dddddddd}
\end{subequations}

In Figs.~\ref{fig:14} (a) and (b), the winding number density \(k_\text{H}\) and vortex number density \(\rho_\text{H}\) are plotted as functions of \(b\) for a fixed value of \(a = 2.5\) along the path \textcircled{7} in Fig.~\ref{fig:phase_figure_wx}. Both quantities decrease as \(b\) increases. When \(b\) is small, the system resides in the perfect vortex lattice phase, i.e., the VL-H phase, and eventually melts into the vortex disorder phase as \(b\) increases.

The type of phase transition can be identified by the signature of the specific heat \(C_\text{V}\). In a BKT phase transition, the specific heat \(C_\text{V}\) typically does not show a singularity as the system approaches the critical region. As the size $L$ increases, the peak of the specific heat converges. In Figs.~\ref{fig:14} (c) and (d), both the specific heat \(C_\text{V}\) and the quantity \(C_\text{H}\) are plotted, and both converge as \(L\) approaches 96 and 128. In addition, both path $\textcircled{6}$ and path $\textcircled{7}$ in Fig.~\ref{fig:phase_figure_wx} exhibit a BKT transition, which follows the same universal behavior and will not be elaborated here. Furthermore, we also measure the correlation function and find that within the VL-H and VL-T phases, the vortex correlation function decays as a power law, indicating the existence of vortex quasi-long-range order, which serves as supplementary evidence for the aforementioned BKT transition.

\subsubsection{The new exponents $y_t=1.340(3)$ between the phases VL-H and VL-T}

We have also discovered that a phase transition can occur between the two vortex phases, i.e., the VL-T and VL-H phases, along the path \textcircled{8} as shown in Fig.~\ref{fig:phase_figure_wx}.
After attempting,  unlike the Ising model's order-disorder phase transition, we could not find a good Binder ratio of vortex density and energy to determine the phase transition point.

The critical exponent \( y_t \) can be fitted using the following finite-size scaling formula~\cite{ising3d}:
\begin{equation}
  C_\text{V}(L) = C_0 + L^{2y_t - d}(m + nL^{y_1}),  
  \label{eq:Cv}
\end{equation}
where \( C_\text{V}(L) \) is the specific-heat  to be fitted, \( L \) is the system size, \( C_0 \), \( m \), and \( n \) are constants, and \( d \) is the spatial dimension.

\begin{figure}[t]   \centering\includegraphics[width = 1\linewidth]{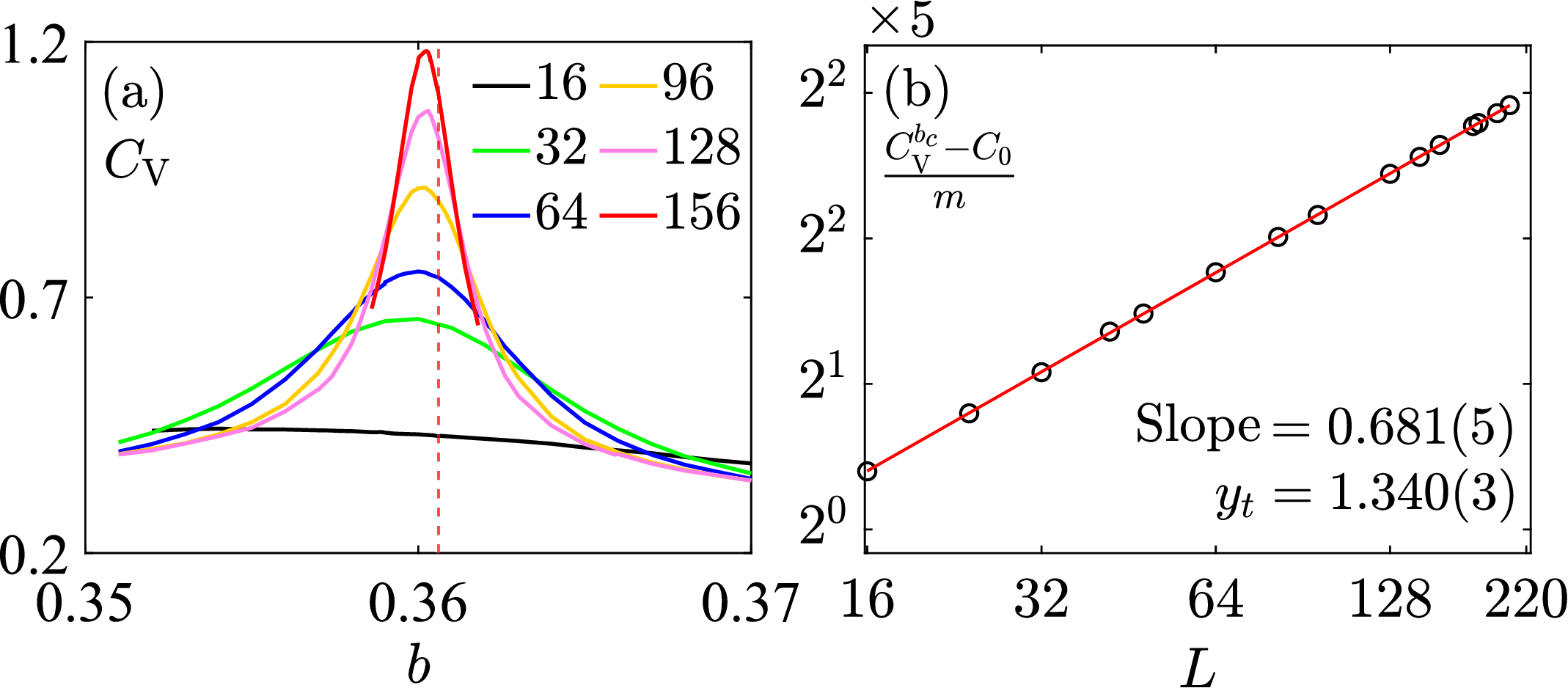}\caption{Finite size scaling of specific heat  (a) The raw data of \( C_\text{V} \) as a function of \( b \) for system sizes \( L = 16 \) to \( L = 156 \).
(b) The data of \( \frac{|C_\text{V}^{b_c} - C_0|}{m} \) versus \( L \) presented in a log-log plot.
}\label{fig:opipo}\end{figure}

In Fig.~\ref{fig:opipo} (a), it contains the raw data of specific heat $C_\text{V}$.
Fortunately, as the size increases from $L$=16 to 156,  the height of the specific heat peak increases sharper. 
It is possible to extrapolate the phase transition point at the thermodynamic limit using the peaks corresponding to their positions $b_c(L)$ of these finite sizes.
 The critical point $b_c(\infty)$=0.3603(2) is obtained using this extrapolation method.
% Figure~\ref{fig:opipo} (b) shows the data of $b_{\text{max}}$ versus $1/L$, where $b_{\text{max}}$ represents the value of $b$ corresponding to the maximum of $C_T$. After discarding the data points for smaller system sizes ($L=16, 32$), we perform an extrapolation to determine the phase transition point, which is found to be $b_c = 0.3605(1)$.

Then, using the values of \( C_\text{V}(L) \) at the critical point \( b_c (L\rightarrow\infty)\) and fitting the data according to Eq.~\ref{eq:Cv}, the following values are obtained: \( 2y_t - d = 0.681(5) \), \( m = 0.030(9) \), and \( C_0 = 0.230(4) \).
In Fig~\ref{fig:opipo} (b),  plotting 
The data points of these two quantities $\frac{C^{b_{c}}_{V}-C_{0}}{m}$ versus $L$ fit the linear behavior very well in log-log plotting.
The conclusion is that the new exponent $y_t=1.340(3)$ is obtained. 

In fact, there is another method to obtain the exponent $y_t$.
 Using  the scaling relationship~\cite{Sandvik2010ComputationalSO},
\begin{equation}
\left|\frac{b_{\text{max}}(L)-b_c(\infty)}{b_c(\infty)}\right| \sim L^{-y_t},
\end{equation}
the exponent $y_t$ can also be obtained.
From the example in this paper, distinguishing a reliable \( b_c(L) \) near the phase transition point (within the range of the fifth decimal place) requires a significant amount of computational effort.
Therefore, the results are not shown here. We leave this task to be completed in the future.

\subsubsection{vortex structural factors} In Ref.~\cite{nature_yw}, the spin and vortex structural factors, defined on the triangular faces, were visualized. As the distances between the magnetic rods increase, which is equivalent to raising the temperature, the peaks of the structural factor gradually disappear, as expected.

\begin{figure}[t]   \centering\includegraphics[width = 1.0\linewidth]{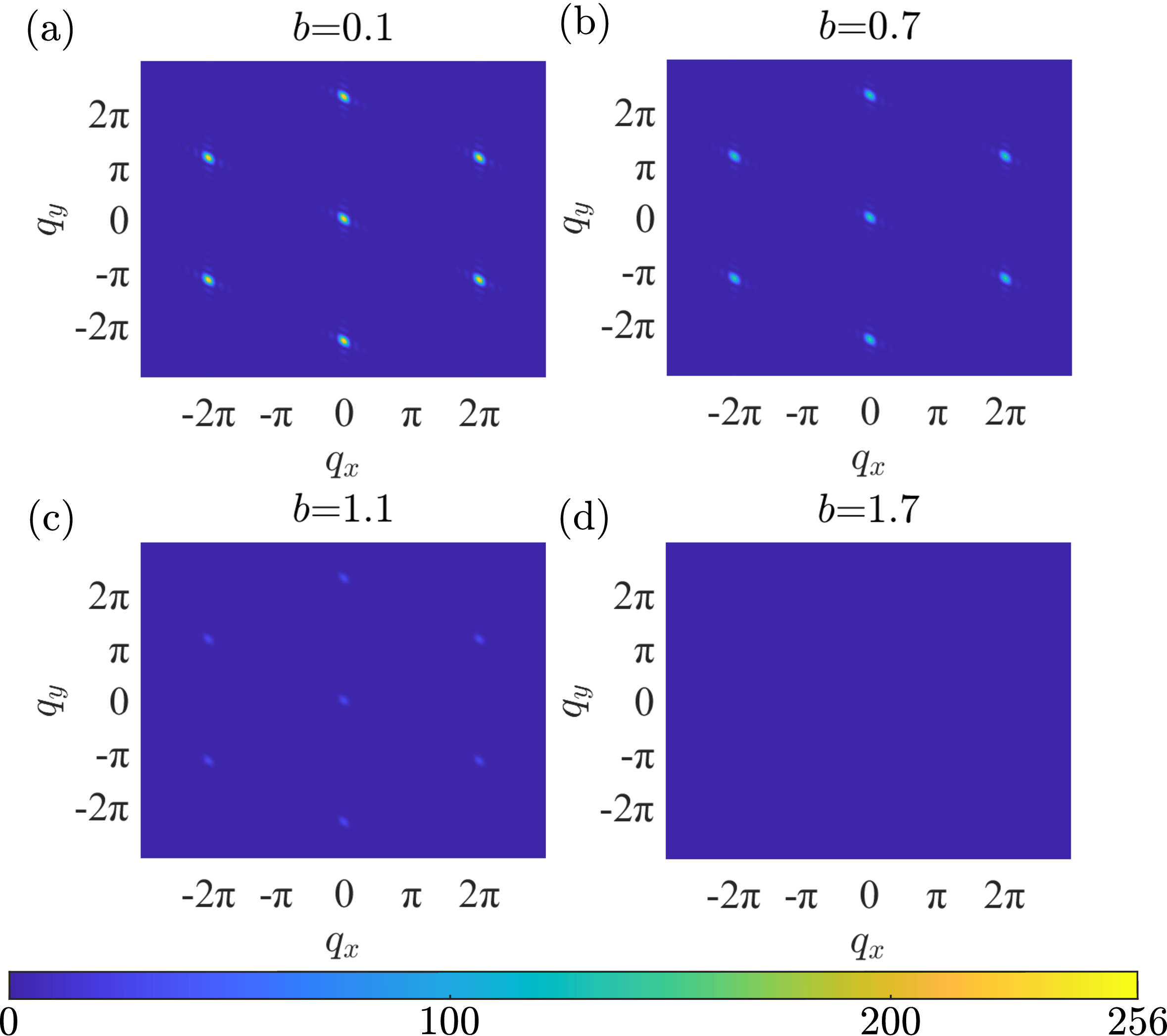}\caption{Schematic representation of the structure factors measured for different values of $b$ when $a = 2.5$.}\label{fig:yyyy}\end{figure}

Here, taking the snapshot in Fig.~\ref{fig:snapshotab} (a) as an example, we define the structure factor of vortex on the honeycomb faces as follows\begin{equation}S(q)=\frac{1}{L^2} \sum_{i,j}^{L^2} e^{iq\cdot(\bm{r_i}-\bm{r_j}) }\left \langle k_ik_j \right \rangle,
\label{eq:sq}\end{equation}
where $L^2$ is number of honeycomb faces. For simplicity,   the distance between the nearest neighboring honeycomb face be 1. $\bm{r_{i}}$ and $ \bm{r_{j}}$ are the true coordinates of the vortex,$k_{i}$ and $k_{j}$ is the value of the number of vortex windings. In Fig.~\ref{fig:yyyy}, the results of \( S(q) \) for a lattice with \( L = 16 \) are shown. Fixing \( a = 2.5 \) and gradually increasing \( b \) from \( b = 0.1 \) to \( b = 0.7 \), \( b = 1.1 \), and \( b = 1.7 \), the peaks progressively darken. This behavior is consistent with the results shown in Fig.~\ref{fig:14}(a), where \( k_\text{H} \) decreases as \( b \) increases.

The location of one of the highlights appears in $(q_{x},q_{y})=(2\pi ,\frac{2\sqrt{3} }{3}\pi  )$. This is due to the fact that the distance between the vortices satisfies the following relationship:\begin{equation}\begin{aligned}\bigtriangleup x&=x_{i}-x_{j}=1,\\\bigtriangleup y&=y_{i}-y_{j}=\sqrt{3}, \\q_{x(y)}&=\frac{2\pi}{\bigtriangleup x(y)},\end{aligned}\end{equation}where the $\bigtriangleup x$ and $\bigtriangleup y$ are marked in red line segments in Fig.~\ref{fig:snapshotab}.

\subsubsection{No fractional vortex phase}Here, we further analyze the four phases from the perspective of vortex excitation. Following~\cite{song1}, the topological charge $q$ in a hexagon is defined as the average value of $k_i$ the six connected triangular faces, as follows\begin{equation}   q=\dfrac{1}{6}\sum_i k_i^\text{T}.   \label{eq:q1}\end{equation} Unlike the pure XY model on the Kagome lattice, the arrows of the six-vertex model on the triangular edges can not form vortex with negative  $k_i^T<0$. This can be confirmed by exhaustively examining the 8 possible states ($+++, \cdots, ---$) of the arrows on the three sides of a triangle. The clockwise and counterclockwise vortex correspond to positive $k_i^\text{T}>0$ or $k_i^\text{T}=0$ for other types of vertex configurations.  

Therefore, in our perfect snapshots, as shown in Figs.~\ref{Fig:snapshot} (a)-(b) and Fig.~\ref{fig:snapshotab} (b), the vortex excitation defined in Eq.~\ref{eq:q1} is  the integer $q=1$ rather than a fraction such as 2/3 or 1/3 in Refs.~\cite{song1,song2} for the XY model.

In addition, on the honeycomb faces, by enumerating all possible configurations, the vortex winding numbers are \( k_\text{H} = -2, -1, 0, 1 \), and there is no case for \( +2 \), still different from the pure XY model on the Kagome lattice.

\subsection{Discussion of possible mechanisms}
~~In paths \textcircled{1}, \textcircled{2} and \textcircled{3} of Fig.~\ref{fig:phase_by_mag}, the phase transition is of the Ising type. The reason for the Ising universality  in the above phase transition can be understood through the $Z_2$ symmetry breaking mechanism. Taking path \textcircled{3}(FM phase to PM phase) as an example, In Fig.~\ref{Fig:snapshot} (b), the degeneracy of the configuration is 2. The energy of the configuration is the same when all arrows are flippe, only the configuration of vertex $b2$ changes to vertex $b1$, and therefore the system exhibits $Z_2$ symmetry breaking. From the vortex perspective at the same location, as $b$ decreases and the system moves from the VL-II phase into the VL-T phase, the arrangement of counterclockwise and clockwise rotations becomes disordered, thus the system exhibits $Z_2$ symmetry.

In paths \textcircled{4} of Fig.~\ref{fig:phase_by_mag}, the phase transition from the PM region to the FM phase does not belong to the Ising class, which is understandable. At this parameter point, since both \( a \) and \( b \) are much larger than 1, the system tends to favor the PM(S)  region. The direction of the arrows only breaks the symmetry in the \( x \) direction, and it is not a perfect ordered-disordered phase transition. In the direction perpendicular to \( x \)-direction, no symmetry breaking occurs. Therefore, it cannot be called a $Z2$ symmetry breaking mechanism.

From the perspective of vortices, when the phase changes from the VL-II phase to the VL-D phase, the vortices on the triangular faces are directly annihilated, rather than swapping the vortices. This may be a new type of phase transition induced by topological excitations or topological annihilation, with a critical exponent of \( y_t = 1.317(6) \).

\begin{figure}[tbh]
   \centering                                    
\includegraphics[width = 0.8\linewidth]{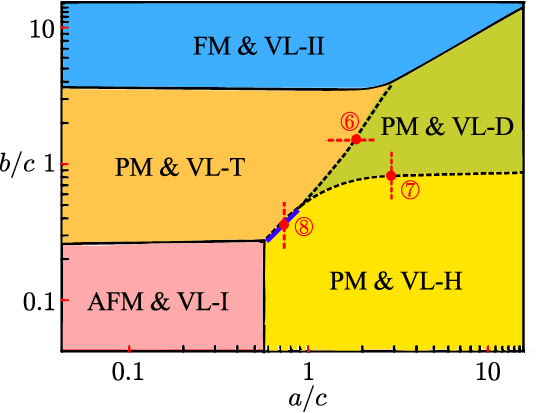}
\caption{The global phase diagram in the plane $a/c-b/c$. The `$\&$' symbol  denotes the simultaneous occurrence of phases determined by magnetization and vortices.}
% For convenience, the labels Fig. 8(a)-Fig. 13(b) are used to mark the positions of the parameters taken for the snapshots drawn later in the phase diagram.}
\label{fig:zongxt}
\end{figure}

\subsection{The total phase-diagram determined by magnetization and vortices}

Taking magnetization and vortex together as the criteria, the phase diagram is summarized as follows: The AFM and VL-I phases occur within the same parameter range, while the FM and VL-II phases appear in the same parameter range. The parameter interval of PM encompasses three vortex phases, namely the VL-T, VL-H, and VL-D phases. For clarity, the complete phase diagram is presented in Fig.~\ref{fig:zongxt}.

\section{Discussion and conclusion}
\label{sec:con}

In this work, we systematically investigated the vortex lattice phases and exponents in Kagome ice using the directed-loop Monte Carlo method. By simulating the theoretical six-vertex model on the Kagome lattice, we identified that 
the phase diagram is asymmetric with respect to \( a = b \), unlike the square lattice, which is symmetric along \( a = b \).

Four distinct VL phases are found.
 Compared to the phases in~\cite{nature_yw}, in our phase diagram, the VL-I phase corresponds to ``Ferrotoroidicity" and  the VL-T phase can be considered analogous to what is described as ``Paratoroidicity", where the vortex density is zero in hexagonal faces and one in triangular faces. However, the other two VL phases and detailed simulations are not reported in Ref.~\cite{nature_yw}. Our findings reveal  the various types of transition between the phases. They are the BKT type between the VL-H(T) and VL-D transition, the Ising type with $y_t=1$ for the FM-PM transiton,  varying exponents $y_t=$1.164(3), 1.317(6), and 1.381(1) for the FM-PM transitions at parameter $a\ge4$,  and $y_t=1.340(3)$ for the VL-T and VL-H transition. These results have not been reported before.
Due to the difficulty in finding an order parameter similar to the Ising model's order-disorder phase transition, the phase transitions between the VL-T and VL-H phases may be related to 
the Landau-incompatible transitions~\cite{PhysRevLett.134.097103}.

Despite these achievements, our work also has room for improvement. First, numerically, the directed loop Monte Carlo method, while effective, may have limitations in exploring more complex models, such as 16-vertex model or disorder vertex model. Advanced computational techniques such as tensor network Monte Carlo~\cite{tnmc2,tnmc} may be required to extend our work.
Second, experimentally, the vertex energy can be directly adjusted in the experiment by altering the positions of the magnetic rods, resulting in 16 different types of vertices~\cite{nature_yw}. However, it remains unclear whether the experiment can restrict the vertex types to the 6V subspace~\cite{nature_yw}, as suggested by the theoretical model.
Our results provide insights into potential realizations in water ice~\cite{dblayer_ice,Hong2024}, colloidal particle-based ice systems~\cite{cj7},  and colloidal active fluids in microchannel networks~\cite{jorge2025active, Jorge2024}.

In summary, our current work contributes to the advancement of numerical methods and experimental techniques to explore interesting phase-transition phenomena in Kagome lattice ice using the simple 6V model.

{\it Acknowledgments} This work  was supported by the Hefei
National Research Center for Physical Sciences at the
Microscale (KF2021002), and the Shanxi Province Science Foundation with Grants No: 202303021221029 (W.Z.) and
No:202103021224051 (J. Z.);
C. D. is supported by the National
Science Foundation of China (NSFC) under Grant Number
11975024; Y. D. was supported by the National Natural Science Foundation of China (Grant
No. 12275263) and the Natural Science Foundation of Fujian Province of China (Grant No. 2023J02032).
\appendix 
\section{Confirm of our code by  enumeration}
\label{sec:a}

\begin{figure}[htbp]
   \centering
\includegraphics[width = 1.0\linewidth]{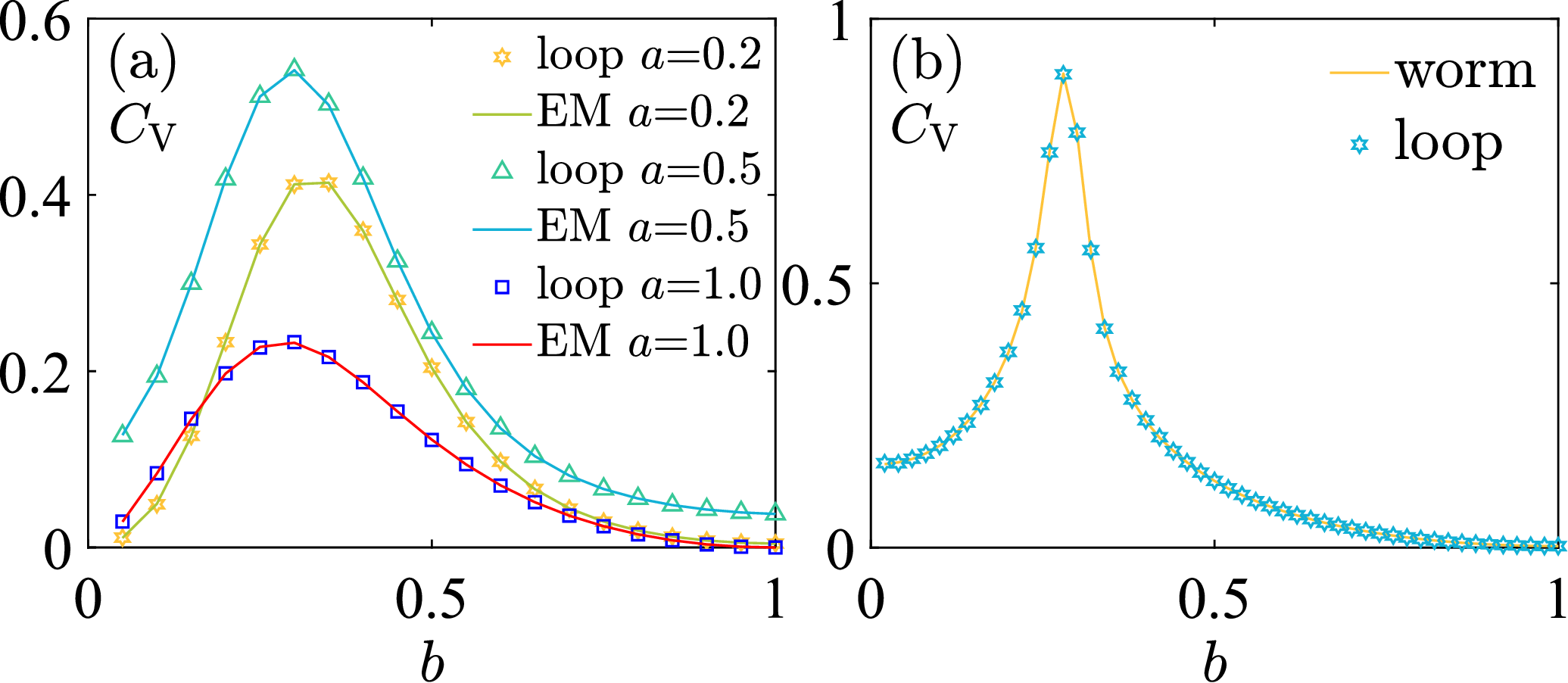}
\caption{Comparison of \(C_\text{V}\) on the \(2 \times 2\) Kagome lattice using the EM and MC methods.(b) With $a=0.5$ and $L=10$, the specific heat are measured by the worm and loop algorithms  and the results are consistent.}
\label{fig:em_vs_mc}
\end{figure}

To ensure the correctness of the code, the results of the directed loop algorithm are compared with those of the enumeration (EM) or brute-force method prior the large-scale simulation. A $2 \times 2$ Kagome lattice is calculated, with 12 vertices ($2 \times 2 \times 3 = 12$) and 24 edges. The results for three different values of $a = 0.2$, 0.5, and 1 are computed while scanning $b$. The results obtained from both methods are the same, as shown in Fig.~\ref{fig:em_vs_mc}.

\section{Worm algorithm}

\label{sec:b}

The following provides the worm algorithm and some results, which are also used to verify the results of the directed loop algorithm.

\begin{enumerate}
\item Pick a random site \( u \) in the Kagome lattice and randomly select one of its neighbors, denoted as \( v \). The spin (or arrows) between \( u \) and \( v \) is flipped with a certain probability \( p_{\text{acc}} \), and as a result, a pair of defects may appear in the Green function space \( G \). Two types of defects are introduced: ``three in and one out" and ``three out and one in“ with the weight of these vertices given by \( d \).
  The acceptance probability can be
\begin{equation}
p_{acc}=\frac{w(u')w(v')}{w(u)w(v)},
\label{eq:pacc}
\end{equation}
where $w(u')$ and $w(v')$ are the probabilities of the type of vertex after flipping the spins. If it is rejected, the round ends, if it is accepted, it enters the $G$-space and there is $(I,M) = (u,v)$ or $(v,u)$.

   \item Randomly pick a neighbor of $I$ called $I'$, try to flip the spin between $I$ and $I'$, if after flipping there is an undefined vertex, then $p_{acc}$=0, otherwise the probability is $p_{acc}$ defined in Eq.~\ref{eq:pacc}. If rejected, repeat the second step, if accepted, move $I $ to $I'$, if $I'$ is $M$, this round of updating ends, otherwise repeat step 2.
\end{enumerate}

\section{An example to calculate the vortex winding number}
\label{sec:c}

\begin{figure}[htbp] % 这里使用figure环境是为了让内容可
\includegraphics[width=0.2\textwidth]{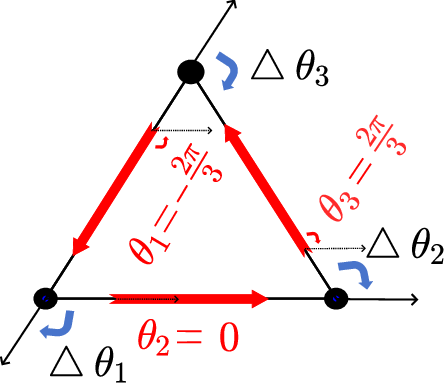} 
    \caption{An example of the winding number calculation of using the directions of a triangle’s three edges.}
    \label{fig:tri}
  \end{figure}

As an illustrative example, we calculate the vortex winding number for a counterclockwise triangle, shown in Fig.~\ref{fig:tri}.
The angles of $\theta_1$, $\theta_2$, and $\theta_3$ are 0, $\frac{2\pi}{3}$, and $-\frac{2\pi}{3}$ respectively.
\quad

The angular differences $\bigtriangleup \theta _{1}$, $\bigtriangleup \theta _{2}$, and $\bigtriangleup \theta _{3}$ are first calculated, with results given by:  
\begin{equation}
\begin{aligned}
\bigtriangleup \theta _{1} &=\theta _{2}-\theta _{1}=0 -\frac{-2\pi}{3} =\frac{2\pi}{3},\\
\bigtriangleup \theta _{2} &=\theta _{3}-\theta _{2}=\frac{2\pi}{3} -0=\frac{2\pi}{3},\\
\bigtriangleup \theta _{3} &=\theta _{1}-\theta _{3}=\frac{-2\pi}{3} -\frac{2\pi}{3}=\frac{-4\pi}{3}.
\end{aligned}
\end{equation}  
% \vskip 1cm

~~\\
Next, to constrain the range of these angular differences, the saw function defined in Eq.~\ref{eq:saw} is applied. The  results are as follows:  
$$ 
\begin{aligned} 
\text{saw}(\Delta\theta_{1}) &= \Delta\theta_{1} = \frac{2\pi}{3}, \\
\text{saw}(\Delta\theta_{2}) &= \Delta\theta_{2} = \frac{2\pi}{3}, \\
\text{saw}(\Delta\theta_{3}) &= \Delta\theta_{3} + 2\pi = \frac{-4\pi}{3} + 2\pi = \frac{2\pi}{3}.
\end{aligned} 
$$  
With the angular differences constrained by the saw function, the vortex winding numbers can be explicitly evaluated using the circulation integral in Eq.~\ref{eq:woxuan}:  
$$ \oint_c \nabla\theta \cdot d\vec{l} = \sum_{i=1}^3 \text{saw}(\Delta\theta_i) = 2\pi \quad \Rightarrow \quad k=+1 .$$ 

% \section{\textcolor{red}{The detail of fitting the exponents 
%  }}
%\label{sec:d}

 \bibliography{main}

\end{document}